\documentclass[%
 reprint,
nofootinbib,
 amsmath,amssymb,
 aps,
]{revtex4-2}

\usepackage{graphicx}
\usepackage{dcolumn}
\usepackage{bm}
\usepackage{hyperref}
\usepackage{makecell}
\usepackage{subfigure}
\usepackage{xcolor}

\begin{document}

\preprint{APS/123-QED}

\title{Photometric redshifts and intrinsic alignments: degeneracies and biases in 3$\times$2pt analysis}

\author{C. Danielle Leonard$^{1}$}
\email{danielle.leonard@newcastle.ac.uk}
\author{Markus Michael Rau$^{1, 2}$}%
\author{Rachel Mandelbaum$^{3}$}

\affiliation{
$^{1}$School of Mathematics, Statistics and Physics, Newcastle University, Newcastle upon Tyne, NE1 7RU, United Kingdom 
}%
\affiliation{
 $^{2}$High Energy Physics Division, Argonne National Laboratory, Lemont, IL 60439, USA
}
\affiliation{
$^{3}$McWilliams Center for Cosmology, Department of Physics, Carnegie Mellon University, Pittsburgh, PA 15213, USA
}

\date{\today}

\begin{abstract}
We present a systematic study of cosmological parameter bias in weak lensing and large-scale structure analyses for upcoming imaging surveys induced by the interplay of intrinsic alignments (IA) and photometric redshift (photo-z) model mis-specification error. We first examine the degeneracies between the parameters of the Tidal Alignment - Tidal Torquing (TATT) model for IA and of a photo-z model including a mean shift ($\Delta \bar{z}$) and variance ($\sigma_{z}$) for each tomographic bin of lenses and sources, under a variety of underlying true IA behaviors. We identify strong degeneracies between: (1) the redshift scaling of the tidal alignment amplitude and the mean shift and variances of source bins, (2) the redshift scaling of the tidal torquing amplitude and the variance of the lowest-$z$ source bin, and (3) the IA source density weighting and the mean shift and variance of several source bins. We then use this information to guide our exploration of the level of cosmological parameter bias which can be induced given incorrect modelling of IA, photo-z, or both. We find that marginalizing over all the parameters of TATT is generally sufficient to preclude cosmological parameter bias in the scenarios we consider. However, this does not necessarily mean that IA and photo-z parameters are themselves unbiased, nor does it mean that the best-fit model is a good fit to the data. We also find scenarios where the inferred parameters produce $\chi^2_{\rm DOF}$ values indicative of a good fit but cosmological parameter bias is significant, particularly when the IA source density weighting parameter is not marginalized over. 
\end{abstract}

\maketitle


\section{Introduction and Motivation}
\label{sec:intro}
The combination of weak gravitational lensing and galaxy clustering measurements produces an extremely powerful probe of cosmological parameters, providing particular insight into the value of $S_8 \equiv \sigma_8 \sqrt{\Omega_{\rm M}/0.3}$, where $\sigma_8$ quantifies the amplitude of matter density fluctuations on a scale of $8\,{\rm Mpc/h}$ and $\Omega_{\rm M}$ is the fractional energy density of matter today \cite{abbott2022dark, heymans2021kids, miyatake2023hyper, sugiyama2023hyper}. Tightening the constraint on $S_8$ is of special interest at the moment, due to the tension which has emerged in recent years \citep{abdalla2022cosmology} between its value as measured by late-time observables vs that implied by measurements from early times (particularly the cosmic microwave background or CMB) \citep{aghanim2020planck, heymans2021kids, abbott2022dark}. Rapidly approaching Stage IV surveys {\it Euclid} \citep{Euclid2022}, the Nancy Grace Roman Space Telescope \citep{spergel2015wide}, and the Rubin Observatory's Legacy Survey of Space and Time (LSST, \cite{Ivezic2019}) will provide an order-of-magnitude increase in our weak lensing source galaxy sample sizes, and will correspondingly produce measurements of weak lensing and galaxy clustering with unprecedented precision. We thus have the opportunity to significantly tighten the constraint on $S_8$. At the same time, these surveys will provide us with uniquely powerful measurements of the equation of state of dark energy, and specifically its parameters $w_0$ and $w_a$, in order to pin down the nature of dark energy as a function of cosmic time. To achieve these goals in a robust manner, though, we must ensure that we strictly control the systematic uncertainties which could impact our analysis. For a review of these systematic effects and their possible impact on cosmological inference in upcoming surveys, see \cite{2018ARA&A..56..393M}.

The canonical analysis set-up for the current and upcoming generation of photometric cosmological galaxy survey samples is the so-called 3$\times$2pt analysis, which consistently combines measurements of weak lensing and galaxy clustering two-point correlation functions in tomographic redshift bins (see, for example, \cite{abbott2022dark, heymans2021kids, prat2022catalog}). Two essential sources of systematic modelling uncertainties for 3$\times$2pt analysis are photometric redshifts (photo-z) and the intrinsic alignment of galaxies (IA). Both of these phenomena have been extensively shown to be individually capable of biasing cosmological parameter constraints when inadequately modelled (see e.g. \cite{kirk2015galaxy, krause2016impact} for IA and \cite{hearin2010general, bernstein2010catastrophic, mandelbaum2018lsst} for photo-z). We have also recently seen that at the level of current data, selecting the appropriate IA model is non-trivial, and opting for a more complex model when a simpler one would suffice has the potential to induce projection effects and act as a limiting factor in 3$\times$2pt cosmological constraints \citep{secco2022dark}.

Typically, tests as to the appropriate model for a given systematic effect are conducted effect-by-effect, fixing the modelling of those not under active consideration (see, for example, \cite{krause2021dark}). However, recent works \citep{fischbacher2023redshift, wright2020kids, stolzner2021self} have begun to evidence a connection between the modelling of IA and the treatment of photometric redshifts, showing that differences in photometric redshift modelling and residual uncertainty in the parameters of such models can lead to biases to the intrinsic alignment amplitude, as well as to cosmological parameters including $S_8$. In particular, \cite{fischbacher2023redshift} showed that for a cosmic-shear-only analysis, mis-specifying the photo-z model, particularly at lower redshifts, can induce a bias in IA amplitudes, and also that the level of IA present in the data impacts the degree to which cosmological constraints are biased due to photometric redshift errors.

In this work, we consider the interplay between intrinsic alignment and photometric redshift model mis-specification in the context of a 3$\times$2pt analysis for a Stage IV survey. Working within the scope of state-of-the-art modelling of these two effects, we identify the most significant degeneracies which occur between photo-z and IA parameters under an expansive set of IA truth scenarios. We find that the nature of the true IA which is present does substantially impact the dominant degeneracies identified. 

Having identified three scenarios in which significant degeneracies occur between IA and photo-z parameters, we explore each case in more detail. Specifically, we consider to what extent the presence of these degeneracies, together with a mis-specification of one or more of the parameters involved, can induce bias in the cosmological parameters of interest. We focus on bias in the 2D planes of $S_8 - \Omega_{\rm M}$ as well as of $w_0 - w_a$. Following the recent works of \citep{campos2022empirical, paopiamsap2023accuracy}, we also consider the effect on the $\chi^2$-per-degree-of-freedom statistic.

In Section~\ref{sec:theory}, we introduce the theoretical framework for 3$\times$2pt analysis, the IA and photo-z models we will consider, and the statistical framework for our parameter inference bias calculations. Then in Section~\ref{sec:finddegs} we first discuss the identification of relevant degeneracies between IA and photo-z parameters under different IA truth models, before moving on in Section~\ref{sec:bias} to explore these degeneracies in details insofar as their potential to induce biases in cosmological parameter inference analysis. We conclude in Section~\ref{sec:conc}.

\section{Theory and Setup}
\label{sec:theory}
In this Section, we review the theoretical formalism for modelling in a 3$\times$2pt analysis (including photo-z and IA modelling). We largely follow the modelling formalism described in \cite{secco2022dark, krause2021dark}. We proceed to review the Fisher forecasting formalism, and introduce our particular forecasting set-up as well as our metrics for parameter bias and goodness-of-fit.

\subsection{Modelling}
\label{subsec:mod}

\subsubsection{3$\times$2pt spectra}
\label{subsubsec:3x2pt}
We consider a so-called 3$\times$2pt set of observables; that is, our observables are the three possible combinations of angular two-point correlation functions between galaxy positions and shears. In Fourier space, where we choose to work, the theoretical expression for the two-point spectrum of galaxy clustering is given by
\begin{align}
C_{g^i g^j}^\ell&=\int_{0}^{\chi_\infty}d\chi b_i(\ell, \chi) b_j(\ell, \chi) \frac{dN^i}{d\chi}\frac{dN^j}{d\chi} P_{\delta}\left(\frac{\ell}{\chi}, \chi \right),
\label{Clgg}
\end{align}
where $i$ and $j$ denote tomographic redshift bins, $b_i$ is galaxy bias (which for generality we write as a function of time and scale), $\frac{dN^i}{d\chi}$ is the distribution of galaxies in tomographic bin $i$ with respect to comoving distance $\chi$, $\chi_\infty$ is the comoving distance to the horizon, $P_\delta$ is the matter power spectrum, and here and throughout unless otherwise noted we assume a flat Universe and employ the Limber approximation \citep{limber1953analysis}.

Similarly, in the case of galaxy-galaxy lensing, the angular two-point spectrum is given by
\begin{equation}
C_{\kappa^i g^j}^\ell= \int_{0}^{\chi_\infty}d\chi b(\ell, \chi) \frac{W_i(\chi)}{\chi}\frac{dN^j}{d\chi} P_{\delta}\left(\frac{\ell}{\chi}, \chi \right),
\label{Clkg}
\end{equation}
where $W_i(\chi)$ is the lensing efficiency for tomographic bin $i$, given by (see e.g. \cite{secco2022dark})
\begin{equation}
W_i(\chi) = \frac{3 H_0^2 \Omega_{\rm M}}{2c^2}\frac{\chi}{a(\chi)} \int_{\chi}^{\chi_\infty} \frac{\chi'-\chi}{\chi'}\frac{dN^i}{d\chi'} d\chi'.
\label{Wdef}
\end{equation}
Here, $H_0$ is the Hubble constant, $c$ is the speed of light in a vacuum, and $a$ is the scale factor. Finally, the angular spectrum for cosmic shear is given by
\begin{equation}
C_{\kappa^i \kappa^j}^\ell= \int_{0}^{\chi_{\infty}}d\chi \frac{W_i(\chi)W_j(\chi)}{\chi^2} P_{\delta}\left(\frac{\ell}{\chi}, \chi \right).
\label{Clkk}
\end{equation}

We use version 2.8.0 of the Core Cosmology Library\footnote{https://github.com/LSSTDESC/CCL} ({\tt CCL}, \cite{chisari2019core}) to calculate these quantities in this work.

\subsubsection{Photometric redshift distribution}
\label{subsec:photoz_theory}
We qualitatively model the photometric redshift error induced mis-specifications on the sample redshift distribution in a given tomographic lens or source bin $i$ using a parametric model. Our method here follows that of \cite{mandelbaum2018lsst}.

We begin with a reference redshift distribution of the full sample, representing either all lenses or all sources as appropriate. In this section, this distribution is referred to as $\frac{dN}{dz}$. We multiply $\frac{dN}{dz}$ by a top-hat selection function, $\Theta^i(z)$, for each tomographic bin, resulting in a redshift distribution for each tomographic bin $i$ which has sharp cuts in $z$ and incorporates no photo-z error effects yet. To incorporate these effects, we convolve each $\frac{dN}{dz}\Theta^i(z)$ with a Gaussian, which applies a possible shift in the mean as well as a redshift-dependent variance. The resulting set of tomographic redshift distributions are denoted as $\frac{dN^i}{d\tilde{z}}$.

Mathematically, we can write this as:
\begin{align}
    \frac{dN^i}{d\tilde{z}}&= \frac{\int dz \frac{dN}{dz}\Theta^i(z) p^i(z, \tilde{z})}{\int d\tilde{z} \int dz \frac{dN}{dz}\Theta^i(z) p^i(z, \tilde{z})} \, .
    \label{dNdzp}
\end{align}
where the denominator is simply a normalization factor. $p^i(z, \tilde{z})$ is the above-mentioned Gaussian and is given by
\begin{align}
    p^i&(z, \tilde{z}) \propto \exp\left(-\frac{1}{2}\left(\frac{z - (\tilde{z} + \Delta \bar{z}^i)}{\sigma_z^i(1+z)}\right)^2 \right)
    \label{pzszp}
\end{align}
where $\Delta \bar{z}^i$ parameterizes the shift in the mean of the tomographic bin and $\sigma_z^i$ the variance imposed by photo-z error. 

Equation~\eqref{dNdzp} is a qualitative model that describes how photometric redshift systematics change the shape of tomographic redshift bins, where $\Delta \bar{z}^i$ and $\sigma_z^i$ represent typical degrees of freedom. Neither the construction of equation \ref{dNdzp}, nor $z$ and $\tilde{z}$ should be associated with realistic sample redshift inference or redshift estimates. We refer to e.g. \cite{myles2021dark, bilicki2018photometric, rau2023weak, schmidt2020evaluation} for more in-depth discussion on sample redshift inference as well as examples of its practical implementation. 

Finally, whilst we acknowledge that a Gaussian-only model such as equation~\eqref{pzszp} is an overly simplistic parameterization of the effect of photo-z error, particularly for the deep source galaxy samples we expect for Stage IV (see e.g. \cite{moskowitz2023improved, stylianou2022sensitivity}), we choose to employ this simplified model in this work as a starting point. Future work involving more complex modelling (e.g. including catastrophic outliers) would no doubt be beneficial, but would itself benefit from building upon the baseline case we present here.

\subsubsection{Intrinsic alignment models}
\label{subsubsec:ia_theory}
\noindent
In equations \ref{Clkg} and \ref{Clkk} above for the two-point functions of galaxy-galaxy lensing and cosmic shear respectively, we have included only the gravitational lensing terms when in reality the measurement of the ensemble shear statistics required to obtain these observables is contaminated by the correlation in the intrinsic alignment of galaxies. For cosmic shear, the more complete two-point function of ellipticities which accounts for this is given by
\begin{align}
C_{\epsilon^i \epsilon^j}^\ell &= C_{\kappa^i \kappa^j}^\ell + C_{\kappa^i I^j}^\ell + C_{I^i \kappa^j}^\ell + C_{ I^i  I^j}^\ell \, ,
\label{IAform}
\end{align}
where $I^i$ is the intrinsic alignment of galaxies in tomographic bin $i$. Similarly, for galaxy-galaxy lensing we have
\begin{align}
C_{\epsilon^i g^j}^\ell &= C_{\kappa^i g^j}^\ell + C_{I^i g^j}^\ell.
\label{IAform_ggl}
\end{align}

It is again helpful to write these angular spectra in terms of projections over the line-of-sight window functions of 3D spectra:
\begin{equation}
C_{\kappa^i I^j}^\ell= \int_{0}^{\chi_\infty}d\chi \frac{W_i(\chi)}{\chi}\frac{dN^j}{d\chi} P_{GI}\left(\frac{\ell}{\chi}, \chi \right),
\label{ClkI}
\end{equation}

\begin{equation}
C_{I^i I^j}^\ell= \int_{0}^{\chi_{\infty}}d\chi \frac{W_i(\chi)W_j(\chi)}{\chi^2} P_{II}\left(\frac{\ell}{\chi}, \chi \right).
\label{ClII}
\end{equation}

\begin{equation}
C_{I^i g^j}^\ell= \int_{0}^{\chi_\infty}d\chi  b_j(\ell, \chi) \frac{dN^i}{d\chi}\frac{dN^j}{d\chi} P_{\delta I}\left(\frac{\ell}{\chi}, \chi \right),
\label{ClIg}
\end{equation}
Our intrinsic alignment model must then provide a means to specify $P_{GI}$, $P_{II}$ and $P_{\delta I}$ in terms of model parameters.

The Tidal Alignment - Tidal Torquing (TATT) model for intrinsic alignment \citep{blazek2019beyond} is a perturbative model which typically takes into account terms up to second-order in the density field and the tidal field (tidal effects being broadly understood to be the main source of IA outside single-halo scales). In the TATT model, the galaxy shape field as sourced by intrinsic alignments is written as:
\begin{equation}
    \gamma^{\rm IA}_{ij} = A_1 s_{ij} + A_{1\delta}\delta s_{ij} + A_2 \sum_k s_{ik}s_{kj} + ...
    \label{gammaTATT}
\end{equation}
where $s_{ij}$ is the tidal field tensor (a position-dependent, 3$\times$3 quantity). $A_1$, $A_{1\delta}$ and $A_2$ are generally $z$ dependent. In accordance with the above expansion, $P_{II}$, $P_{GI}$ and $P_{\delta I}$ can be written in terms of $A_1$, $A_{1\delta}$ and $A_2$; we do not reproduce the full expressions for these here but refer the reader to \cite{blazek2019beyond}. We calculate these 3D spectra and include them in the required projected angular spectra again using {\tt CCL} and specifically the capabilities of the {\tt FASTPT} code incorporated therein \citep{mcewen2016fast}.

We follow \cite{secco2022dark} in parameterizing $A_1$, $A_{1\delta}$ and $A_2$ as follows:
\begin{equation}
    A_1(z) = -a_1 \bar{C}_1 \frac{\rho_{\rm crit}\Omega_{\rm M}}{D(z)}\left(\frac{1+z}{1+z_0}\right)^{\eta_1}
    \label{eq:A1}
\end{equation}
\begin{equation}
    A_2(z) = 5 a_2 \bar{C}_1 \frac{\rho_{\rm crit}\Omega_{\rm M}}{D(z)^2}\left(\frac{1+z}{1+z_0}\right)^{\eta_2}
    \label{eq:A2}
\end{equation}
\begin{equation}
    A_{1\delta}(z) = b_{TA}A_1(z)
    \label{eq:bTA}
\end{equation}
where $\bar{C}_1=5 \times 10^{-14} M_\odot h^{-2}{\rm Mpc}^{2}$ is a normalization factor originally from \cite{brown2002measurement}, $D(z)$ is the linear growth factor, $\rho_{\rm crit}$ is the critical density, and $z_0$ is a pivot redshift which we again follow \cite{secco2022dark} in fixing to 0.62. We see that in this parameterization, the TATT model has 5 free parameters: $\{a_1, \eta_1, a_2, \eta_2, b_{TA}\}$.

In the below, we will also consider the nonlinear alignment model (NLA) \citep{Bridle2007}. The NLA model can be thought of as a reduced version of TATT which considers only the first tidal alignment term; it is equivalent to the TATT model in the case where $a_2=0$ and $b_{\rm TA}=0$ (and $\eta_2$ does not contribute where $a_2=0$). In this work we will consider two versions of the NLA model: one which has two parameters, $a_1$ and $\eta_1$, thus allowing for a $z$-dependent amplitude (which we sometimes refer to as NLA-z below), and one where $\eta_1$ is fixed to 0, forcing a constant intrinsic alignment amplitude with redshift. 

\subsection{Fisher forecasting and bias formalism}
\label{subsubsec:fisher:method}
\noindent
We use Fisher forecasting in this work. The Fisher matrix measures how much information one can extract on a given parameter using a data vector. It is the expectation value of the observed information and, under certain regularity conditions, the upper bound on the parameter precision (the {\it Cramer-Rao Bound}). The components of the Fisher matrix in our scenario are given by:
\begin{equation}
    F_{ij} = \left(\frac{\partial \vec{d}}{\partial \theta_i}\right)^{T} {\rm Cov}^{-1} \frac{\partial \vec{d}}{\partial \theta_j}
\end{equation}
where $\vec{d}$ is the data vector, $\theta_i$ is the $i$-th model parameter, ${\rm Cov}$ is the covariance matrix of the data vector, and derivatives should be understood to be evaluated at a fiducial set of parameter values. The Fisher matrix $F$ then provides a lower bound of the inverse covariance matrix of the estimated model parameter vector $\vec{\theta}$. Note that it has been assumed here that the dependence of the covariance matrix on the parameters $\vec{\theta}$ is negligible.  For a detailed pedagogical introduction to Fisher forecasting, see \citet{bassett2011fisher, blanchard2020euclid}.

In this work, we use the Fisher matrix not only to obtain forecast parameter constraints and degeneracies, but also to estimate the level of bias which an incorrect modelling assumption may induce in our inferred parameter values. The forecast bias to  the inferred value of parameter $\theta_i$ in this case can be approximated as \citep{amara2008systematic, huterer2005calibrating, taylor2007probing}:
\begin{align}
    \delta \theta_i = \left(F_{ij}\right)^{-1}\left(\vec{d}_{\rm true} - \vec{d}_{\rm fid}\right) {\rm Cov}^{-1} \frac{\partial \vec{d}_{\rm fid}}{\partial \theta_j}
    \label{biasF}
\end{align}
where $\vec{d}_{\rm true}$ is the data vector as computed under the `true', correct model and $\vec{d}_{\rm fid}$ is the data vector computed at the fiducial values of the inappropriate (mis-specified) model. Note that $F_{ij}$ is the Fisher matrix within the {\it inappropriate} model i.e. for the parameter set of the model which would be assumed in the hypothetical analysis. 

A common pitfall in Fisher forecasting is the need for numerical derivatives of the data vector with respect to parameters. In this work, we make use of a 5-point stencil method for numerical derivatives. For the results shown in Section~\ref{sec:finddegs} below, we have verified in all cases that results are insensitive to the step size chosen and particularly that there are no qualitative changes due to varying the numerical differentiation step size. Nevertheless, we caution that Fisher forecasting with numerical derivatives almost inevitably carries some level of uncertainty (see, e.g., \cite{bhandari2021fisher}). In Section~\ref{sec:bias}, we compute all results across a range of derivative step sizes, and include error bars which reflect the resulting variation in computed quantities where appropriate. Since our goal is to identify qualitative features of the parameter space and analysis rather than to make quantitative forecasting statements, this derivative-related uncertainty, once considered, does not impeded our work.

Finally, Fisher forecasting intrinsically assumes that the posterior probability distribution of the parameters can be described by a multi-dimensional Gaussian, an assumption which is not always correct but is a suitable approximation in many cases. There is a particular question which may arise about the use of Fisher forecasting in this case, namely the fact that the posterior distributions of the $a_1$ and $a_2$ parameters have been seen in previous analyses to be bimodal, particularly across their 0-values and as a result of the fact that terms scaling like $a_1^2$, $a_2^2$, and $a_1a_2$ are insensitive to a flip in the sign of both parameters. The authors of \cite{secco2022dark} point this out but also note that this is believed to be a feature of the fact that in the current Stage III context, the uncertainties on IA parameters far exceed the magnitude of their best-fit values. We do not expect this to be the case in the Stage IV scenario which we consider here (and the results of our forecasts support this). We additionally mitigate this potential issue via the consideration of a set of truth IA scenarios which encapsulates symmetric cases of both positive and negative $a_1$ and $a_2$ values, thus avoiding the situation in which the one set of potentially equally viable values is not considered. This all being said, Fisher forecasting is ultimately an analysis choice taken for expediency and to minimize computational cost, and future work which employs more realistic analysis methods such as MCMC and variants may be helpful to better quantitatively understand the largely qualitative results expressed here.

\subsection{Forecasting set-up}
\label{subsec:setup}

\subsubsection{Terminology}
\label{subsubsec:terms}
We first clarify the specific meaning we intend for some terms that we will use in describing the forecasts we perform, particularly with regards to the use of different models and modelling choices.
\begin{itemize}
    \item{The {\bf truth model} and {\bf truth parameters} are the models and parameter sets which we use to compute the data vector which acts in the place of the measurement. ${\vec{d}_{\rm true}}$ is this data vector. The truth model and truth parameters are a full and correct description of the underlying physics as presumed in a given forecasting scenario. The truth model is also used in the calculation of the $\Delta (S_8 - \Omega_{\rm M})$ and $\Delta (w_0 - w_a)$ metrics (defined below) as the fraction of $\sigma$ bias is reported with reference to the posterior distribution {\it in the truth model}, following the procedure of \cite{campos2022empirical}.}
    \item{The {\bf assumed model} is the model which would be used to model the data vector in a real analysis. In a real analysis, it would be used in computing the theoretical version of the data vector at a given set of sampled parameters as part of Bayesian parameter inference.}
    \item{The {\bf fiducial parameters} are the parameters of the assumed model to be utilized in computing ${\vec{d}_{\rm fid}}$ and the Fisher matrix in equation~\eqref{biasF}.}
\end{itemize}

\subsubsection{Assumed survey configuration}
\label{subsubsec:survey}
\begin{figure*}
\centering
\subfigure{\includegraphics[width=0.45\textwidth]{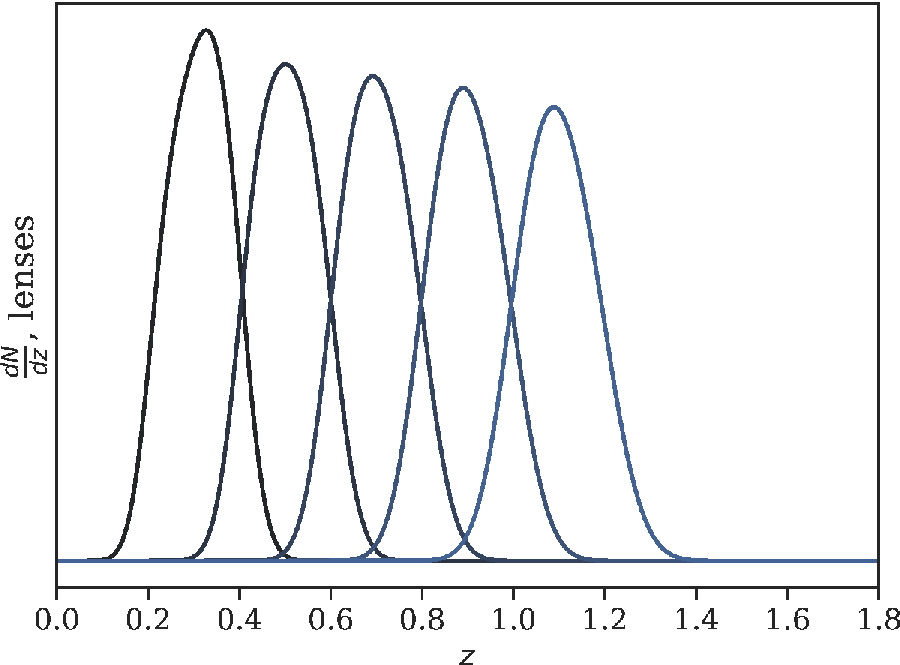}}
\subfigure{\includegraphics[width=0.45\textwidth]{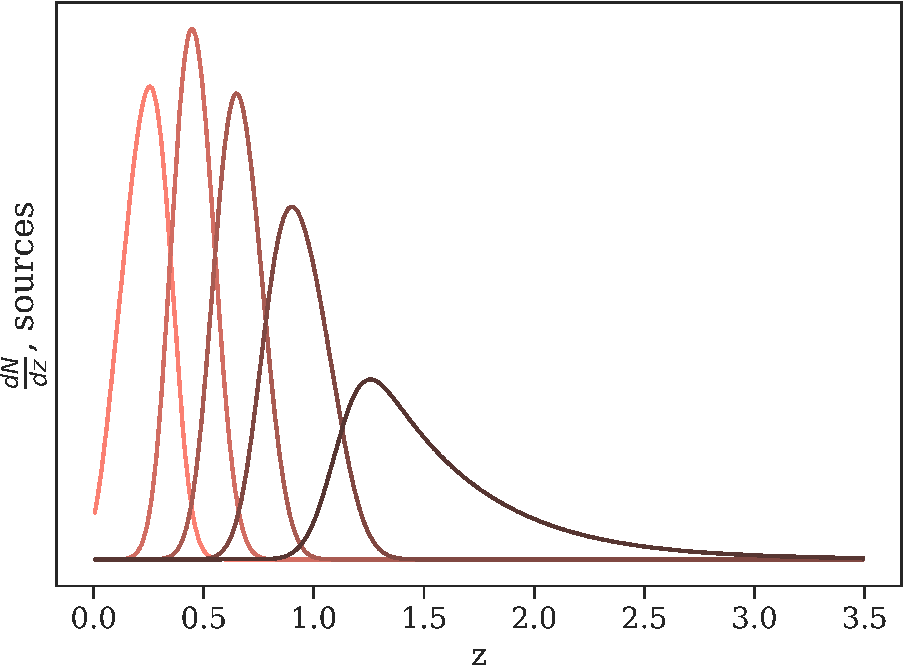}}
\caption{Fiducial lens galaxy (left) and source galaxy (right) redshift distributions in our forecasting analysis set-up.}
\label{fig:dndzs}
\end{figure*}

Our 3$\times$2pt data vector $\vec{d}$ comprises the clustering, galaxy-galaxy lensing, and cosmic shear $C_\ell$'s as given in equations~\eqref{Clgg}, \eqref{Clkg} and \eqref{Clkk} respectively. In an effort to base our observational set-up on a realistic early Stage IV 3$\times$2pt analysis, we largely follow the observational set-up for LSST Year 1 as given in \cite{mandelbaum2018lsst}. This includes:
\begin{itemize}
    \item{5 source redshift bins and 5 lens redshift bins, following the prescription for LSST Year 1 of \cite{mandelbaum2018lsst}. Notably this includes the joint probability of photo-z and true-z given in equation~\eqref{pzszp} above. These redshift distributions are visualized in Figure~\ref{fig:dndzs}}.
    \item{Auto-spectra only for clustering. Galaxy-galaxy lensing spectra selected for inclusion via the same criteria as in \cite{mandelbaum2018lsst} such that cross-correlations where source redshifts are predominantly below lens redshifts are excluded.}
    \item{20 $\ell$ bins for each cross-spectrum, logarithmically spaced between $\ell=20$--15,000.}
    \item{Scale cuts to remove high-$\ell$ (small scale) bins as described in \cite{mandelbaum2018lsst}. For clustering and galaxy-galaxy lensing, the maximum $\ell$ value retained for each spectrum is defined as
    \begin{equation}
        \ell_{\max} = k_{\rm max} \chi(\langle z \rangle) - 0.5
    \end{equation}
    where $k_{\rm max}$ is taken to be 0.3 $h$/Mpc and $\langle z \rangle$ is the mean redshift of the lens bin in question. For cosmic shear, we take $\ell_{\rm max}=3000$ in all cases.
    }
\end{itemize}
We use the data products and scripts released with \cite{mandelbaum2018lsst} to obtain the appropriate LSST Year~1 3$\times$2pt data covariance matrix for our set-up. The data covariance includes non-Gaussian effects and was originally computed using \texttt{CosmoLike}\footnote{https://github.com/CosmoLike/DESC\_SRD} \citep{krause2017cosmolike}. It assumes an LSST observing strategy which results in a survey area of $1.23\times 10^4$ deg$^2$ for Year 1, corresponding to a sky fraction of approximately 0.3.

\subsubsection{Parameter space}
\label{subsubsec:parspace}
Our parameter space consists of cosmological, galaxy bias, intrinsic alignment, and photometric-redshift parameters.

Our cosmological parameter space is summarized in Table \ref{tab:cosm_b_vals}. We impose broad Gaussian priors on the cosmological parameters which are intended to be uninformative but to stabilize the numerical calculation of the Fisher matrix. We then further impose priors which represent Stage-III-like constraints on the subset of cosmological parameter space which excludes $w_0$ and $w_a$, including realistic correlations. We take the Fisher matrices representing these Stage III priors from the data release of \cite{mandelbaum2018lsst}. Note that in the below analysis we make use of the derived parameter $S_8= \sigma_8 \sqrt{\Omega_{\rm M}/0.3}$; see Section 5 of \cite{coe2009fisher} for the procedure by which we transform Fisher matrices to this basis (as well as for a helpful review of Fisher forecasting in general). 

\begin{table*}
\begin{center}
\begin{tabular}{ |c|c|c|c|c|c|c|c|c|c|c|c|c| }
 \hline
&$\Omega_{\rm M}$ & $\Omega_{\rm B}$ & $h$ &$w_0$ & $w_a$ & $\sigma_8$ & $n_s$ & $b_1$ & $b_2$ & $b_3$ & $b_4$ & $b_5$ \\
 \hline
$p_{\rm fid}$ & 0.3156 & 0.0492 & 0.6727 & -1 & 0.05 & 0.83 & 0.9645 & 1.562362 & 1.732963 & 1.913252 & 2.100644 & 2.293210\\
\hline
$\sigma(p)$ & 0.1 & 0.04 & 0.1 & 1 & 5 & 0.14 & 0.1 & - & - & - & - & - \\
 \hline
\end{tabular}
 \caption{Fiducial values, $p_{\rm fid}$, and Gaussian prior widths, $\sigma(p)$, where relevant, of cosmological and galaxy bias parameters $p$. See text for note on additional Stage III cosmological parameter priors. Galaxy bias parameters have uniform priors and their fiducial values are taken in accordance with \protect\cite{mandelbaum2018lsst}.}
\label{tab:cosm_b_vals}
 \end{center}
\end{table*}

We include a scale-independent galaxy bias in each tomographic bin (setting $b(\ell,z)$ in equations~\ref{Clgg} and~\ref{Clkg} above to a constant for each lens galaxy bin). Fiducial values of these bias parameters are taken in agreement with \cite{mandelbaum2018lsst} and given in Table \ref{tab:cosm_b_vals}.

For the parameters which describe intrinsic alignment, several fiducial sets of parameters are required to describe the different model choices we will explore. For intrinsic alignment all models considered are subsets of the TATT model. The sets of fiducial values for the intrinsic alignment models considered below are given in Table~\ref{tab:IA_model_vary}. Priors on IA parameters are uniform and flat.

\begin{table}
\begin{center}
\begin{tabular}{ |c|c|c|c|c|c|  }
 \hline
Par set & $a_1$ & $\eta_1$ & $a_2$ & $\eta_2$ & $b_{\rm TA}$ \\
 \hline
1  &  0.5  & 1.5 & 0.5 & 1.5 & 1\\
2 & 0.5  & 0 & 0.5 & 0 & 1  \\
3 & 0.5  & -1.5 & 0.5& -1.5 & 1  \\
4 & 0.5  & 1.5 & 0 & 0 & 0  \\
5 & 0.5  & 0 & 0 & 0 & 0  \\
6 & 0.5  & -1.5 & 0 & 0 & 0  \\
7 & 0.5  & 1.5 & -0.5 & 1.5 & 1  \\
8 & 0.5  & 0 & -0.5 & 0 & 0  \\
9 & 0.5  & -1.5 & -0.5 & -1.5 & 1  \\
10 & -0.5  & 1.5 & 0.5 & 1.5 & 1  \\
11 & -0.5  & 0 & 0.5 & 0 & 1  \\
12 & -0.5  & -1.5 & 0.5 & -1.5 & 1  \\
13 & -0.5  & 1.5 & 0 & 0 & 0  \\
14 & -0.5  & 0 & 0 & 0 & 0  \\
15 & -0.5  & -1.5 & 0 & 0 & 0  \\
16 & -0.5  & 1.5 & -0.5 & 1.5 & 1  \\
17 & -0.5  & 0 & -0.5 & 0 & 1  \\
 \hline
\end{tabular}
 \caption{Intrinsic alignment parameter sets used in identifying degeneracy directions and exploring parameter inference bias. Note that parameters listed as having fiducial value `0' were in reality given a very small fiducial value of 0.01 in order to prevent numerical issues related to the Fisher analysis framework.}
  \label{tab:IA_model_vary}
 \end{center}
\end{table}

For the modelling of photo-z error, the full set of possible modelling parameters is $\Delta \bar{z}^{i=1-5}_l$, $\sigma_{z,l}^{i=1-5}$, $\Delta \bar{z}_s^{j=1-5}$ and $\sigma_{z,s}^{j=1-5}$, with $l$ indicating lens tomographic bins and $s$ indicating source tomographic bins. We define a `baseline' set of photo-z parameter values, which are provided in Table~\ref{tab:pz_vals}. Priors on photo-z parameters are uniform flat. Note that we will often below consider the case where $\sigma_{z,l}^{i=1-5}$ and $\sigma_{z,s}^{j=1-5}$ must be specified as part of the model, but should be considered as fixed parameters and not ones which would be marginalized over in a parameter inference analysis. We refer to this as the `binned shift' photo-z model. 

\begin{table}
\begin{center}
\begin{tabular}{ |c|c|c|c|  }
 \hline
$\Delta \bar{z}^{i=1-5}_l$ & $\Delta \bar{z}_s^{j=1-5}$ & $\sigma_{z,l}^{i=1-5}$ &$\sigma_{z,s}^{j=1-5}$ \\
 \hline
0  & 0 & 0.05 & 0.03  \\
 \hline
\end{tabular}
 \caption{`Baseline' photo-z parameter values.}
  \label{tab:pz_vals}
 \end{center}
\end{table}

\subsubsection{Metrics}
\label{subsubsec:metrics}
We use three metrics to describe the severity of the cosmological parameter biases and impact on goodness-of-fit that can result from mis-specifying a component of our model. The first two are related to the level of cosmological parameter bias induced by model mis-specification:
\begin{itemize}
    \item{{\bf $\Delta (S_8 - \Omega_{\rm M})$}: This is the absolute level of the parameter bias in the plane of $S_8$ and $\Omega_{\rm M}$ (marginalizing over all other parameters). We report this as $n\sigma$, calculating $n$ by finding the confidence region under an analysis which assumes the truth model on the border of which the biased values of $S_8$ and $\Omega_{\rm M}$ lie. This procedure follows \cite{campos2022empirical}. It is typical in checking for robustness to systematics modelling choices to require that a given modelling choice introduce no more than $0.3 \sigma$ cosmological parameter bias (see e.g. \cite{krause2021dark}), and we will take this value as a subjective threshold for concern.}
    \item{{\bf $\Delta (w_0 - w_a)$}: The same as above, but considering the parameter bias in the plane of $w_0$ and $w_a$, as these are key target parameters for Stage IV dark energy surveys.}

\end{itemize}

The third metric is related to goodness-of-fit. If a modelling choice results in a high bias in cosmological parameters but fails a typical `goodness-of-fit' test, the modelling issue would in principle be identifiable and correctable. We therefore also want to consider some goodness-of-fit metric, to determine the most troubling cases: where cosmological parameter bias exists but the modelling provides a seemingly acceptable fit to the data. For this purpose, we select $\chi^2_{\rm DOF}$, the $\chi^2$ per degree of freedom. Recall that $\chi^2$ is given by:
    \begin{equation}
        \chi^2 = \left(\vec{d}_{\rm fit} - \vec{d}_{\rm obs}\right)^{\rm T}\cdot{\rm Cov}^{-1}\cdot \left(\vec{d}_{\rm fit} - \vec{d}_{\rm obs}\right)
        \label{chi2}
    \end{equation}
    where $\vec{d}_{\rm fit}$ is the theoretical data vector described by the best-fit model and $\vec{d}_{\rm obs}$ is the observed data vector. $\chi^2_{\rm DOF}$ is then given by $\chi^2 / n_{\rm DOF}$ where $n_{\rm DOF}$ is the number of degrees of freedom: the number of elements of the data vector minus the number of parameters in the model being fit, in the case where all parameters have uninformative priors. Note that because we include informative Stage III priors on some of our cosmological parameters, this definition slightly overestimates our number of degrees of freedom. However, because in our case $n_{\rm DOF}$ is heavily dominated by the number of elements in the data vector, we do not expect this to impact our results.

    An optimal fit would result in $\chi^2_{\rm DOF} \approx 1$, and a `good fit' would thus have $\chi^2_{\rm DOF}$ `near to 1' in some sense. The probability-to-exceed (`p value') quantifies this, where a smaller p-value indicates a worse goodness of fit. For example, p=0.05 indicates that, in the case where the model is a good fit to the data, for 95\% of random draws from the data distribution the $\chi^2$ metric would have been smaller. We thus want to set a (subjective) threshold in p-value for which we consider the fit to be acceptable. Our $n_{\rm DOF}$ ranges from 516-520 (ignoring informative prior effects); for this case, a $\chi^2_{\rm DOF} \approx 1.1$ is equivalent to p=0.05, and we take this as a threshold at which, using this metric, we would be able to empirically identify a potential model mis-specification.

     We create 200 realizations of $\vec{d}_{\rm obs}$ by drawing from a multivariate Gaussian with mean $\vec{d}_{\rm true}$ (the theoretical data vector computed in the truth model) and covariance ${\rm Cov}$. We then compute $\chi^2 / n_{\rm DOF}$ for each of these realizations with $\vec{d}_{\rm fit}$ being given by the theoretical model computed at the biased parameter values as would be inferred under model mis-specification (see equation~\ref{biasF}). In Section~\ref{sec:bias} below, values of $\chi^2_{\rm DOF}$ shown are given as the mean of this set, and error bars on these values are given by the variance of the set. 

    We select $\chi^2_{\rm DOF}$ as our goodness-of-fit metric because of its widespread use in late-time observational cosmology as a heuristic goodness-of-fit diagnostic. We consider it to be the most universally employed metric and hence the one most relevant to this discussion. However, we note there exist a variety of goodness-of-fit and model-selection metrics which have been applied to cosmological scenarios, such as the Akaike Information Criteria \citep{cavanaugh2019akaike}, the Bayesian Information Criteria \citep{knuth2015bayesian}, and the Bayesian evidence \citep{neath2012bayesian}.

When presenting results in terms of all three of these metrics below, we include uncertainties. We stress that the uncertainties presented on $\Delta (S_8 - \Omega_{\rm M})$ and on $\Delta (w_0 - w_a)$ are sourced from the propagation of numerical differentiation errors typical of Fisher forecasting (discussed above in Section~\ref{subsubsec:fisher:method}). The uncertainties presented on $\chi^2_{\rm DOF}$, on the other hand, represent the distribution of $\chi^2_{\rm DOF}$ values as corresponding to a noisy data vector. The effect of numerical differentiation error on the mean $\chi^2_{\rm DOF}$ was investigated and found to be negligible.

\section{IA and Photo-z degeneracies}
\label{sec:finddegs}
Having established our set-up and theoretical framework, we now seek to identify instances of strong degeneracy between photo-z and IA parameters. Such degeneracies indicate cases in which IA and photo-z parameters have highly correlated impacts on the data vector, meaning that in the absence of external priors, their values will normally be poorly constrained. In addition, these instances of parameter correlations raise the possibility of `trade-offs' between IA and photo-z parameter values which may propagate into cosmological parameter bias; we will investigate this possibility further in Section~\ref{sec:bias}. 

For the purpose of this exercise of degeneracy identification, we work under the maximally expansive assumed set of models for IA and photo-z within our framework: TATT for IA and the binned shift + variance model for photo-z. Fiducial cosmological and galaxy bias parameters are set as in Table \ref{tab:cosm_b_vals}, and fiducial photo-z parameters are the baseline set of Table~\ref{tab:pz_vals}. For the fiducial IA parameters, we explore a variety of choices, as current physical priors which could be applied to TATT parameters are broad and their fiducial values have the potential to significantly impact the parameter degeneracies we find \citep{samuroff2019dark, secco2022dark}. We consider 18 different fiducial IA parameter sets, designed to cover qualitatively different situations within the assumed TATT model; these are provided in Table \ref{tab:IA_model_vary}. Note that we vary all 5 TATT parameters in each case, even when a parameter is fiducially set to be negligible.

\begin{figure*}
\centering
\subfigure{\includegraphics[width=0.95\textwidth]{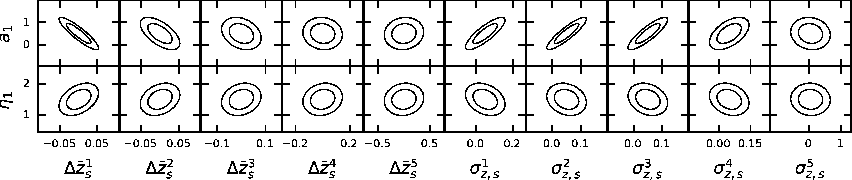}}
\caption{Degeneracies between redshift-dependent NLA parameters $a_1$, $\eta_1$  and source  photo-z parameters $\bar{z}_s$, $\sigma_{z,s}$ under IA parameter set 4 (where $a_2$, $\eta_2$, $b_{\rm TA}$ are all set to 0).}
\label{fig:a1eta1_zssigzs}
\end{figure*}

\begin{figure*}
\centering
\subfigure{\includegraphics[width=0.8\textwidth]{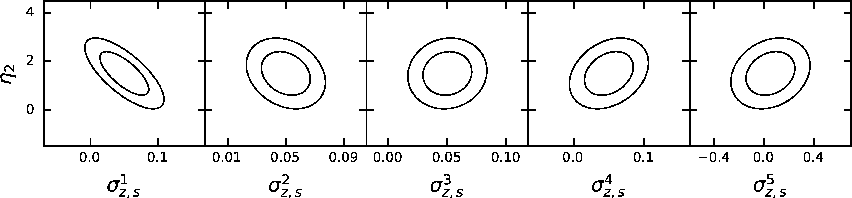}}
\caption{Degeneracy between TATT parameter $\eta_2$ and photo-z parameters $\sigma_{z,s}$ under IA parameter set 1 (which has $a_2$, $\eta_2$, and $b_{\rm TA}$ all non-zero and positive).}
\label{fig:degfind1}
\end{figure*}

\begin{figure*}
\centering
\subfigure{\includegraphics[width=0.5\textwidth]{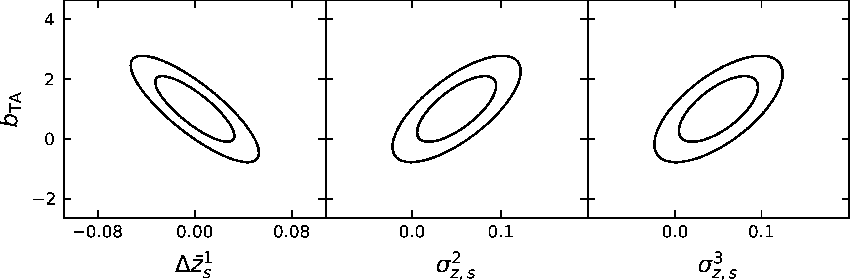}}
\caption{Degeneracy between TATT parameter $b_{\rm TA}$ and photo-z parameters $\bar{z}_s^1$, $\sigma_{z,s}^2$ and $\sigma_{z,s}^3$ under IA parameter set 10 (which has all 5 TATT parameters non-zero, $a_1$ negative and the remainder positive).}
\label{fig:degbTAzssigzs}
\end{figure*}

To identify significant degeneracies, we apply the following procedure, repeating for each set of IA parameters:
\begin{itemize}
    \item{Compute the Fisher matrix.}
    \item{Invert to get the parameter covariance matrix (`Cov').}
    \item{Apply
    \begin{align}
        \,\,\,\,\,\,\,\,\,\,\,\,\,\, {\rm Corr}_{ij} &= \frac{{\rm Cov}_{ij}}{\sqrt{{\rm Cov}_{ii}{\rm Cov}_{jj}}}
    \end{align}
    to get the parameter correlation matrix (`Corr').
    }
    \item{Identify the cases where the cross-correlation coefficient between parameters is relatively large in absolute value. As a rule of thumb, we choose to consider cases where $|{\rm Corr}_{ij}| \geq 0.6$ as cases of interest, however there is clearly variation in strength of degeneracy within those instances above this threshold and this threshold value is somewhat arbitrary. This is not a major issue for our purposes as we are seeking to identify qualitative trends.} 
\end{itemize}
This process reveals the following cases of strong degeneracy between IA and photo-z parameters:
\begin{itemize}
    \item{$a_1$ and $\eta_1$ frequently both display strong degeneracy with both $\Delta \bar{z}_s^i$ and $\sigma_{z,s}^i$ for some subset of redshift bins. Although these degeneracies are evident at some level regardless of the fiducial IA parameter values chosen, they are strongest (often with correlation coefficients $\geq 0.8$) when the fiducial IA parameter values have $a_2=b_{\rm TA}=0$ (a fiducial IA model of redshift-dependent NLA). In other words, the case where the tidal alignment term maximally dominates over tidal torquing maximizes the strength of the degeneracy between photo-z parameters and the tidal alignment parameters $a_1$ and $\eta_1$. We illustrate this in Figure \ref{fig:a1eta1_zssigzs} for the case of IA fiducial parameter set 4; Table \ref{tab:corr_coeffs_a1eta1} displays the correlation coefficients for this case between those IA and photo-z parameters with significant degeneracies.} 
    \item{A strong degeneracy is present between $\eta_2$ and $\sigma_{z,s}^1$ in cases where the fiducial IA parameter has $a_2$ (and also $b_{\rm TA}$) non-zero. For example, for fiducial IA parameter set 1, the correlation coefficient between $\eta_2$ and $\sigma_{z,s}^1$ is -0.73. We can think of this as the tidal-torquing analogue to the above: when tidal torquing is fiducially non-zero, we see significant degeneracy between the parameters of this IA term and the photo-z parameters. This is illustrated in Figure \ref{fig:degfind1} for IA parameter set 1.} 
    \item{There are some instances in which a strong degeneracy is present between $b_{\rm TA}$ and $\Delta \bar{z}_s^1$ as well as between $b_{\rm TA}$ and $\sigma_{z,s}^i$ for $i=2-3$. These degeneracies appear in cases where the fiducial value of $b_{\rm TA}$ is non-zero and are most prominent when $a_1$<0. We see an example in Figure \ref{fig:degbTAzssigzs} for fiducial IA parameter set 10. In this case, $b_{\rm TA}$ has a correlation coefficient of -0.83, 0.69 and 0.64 with photo-z parameters $\Delta \bar{z}_s^1$, $\sigma_{z,s}^2$, and $\sigma_{z,s}^3$ respectively.}
\end{itemize}

\begin{table}
\begin{center}
\begin{tabular}{ |c|c|c|c|c| }
 \hline
  & $\Delta \bar{z}_s^1$ & $\sigma_{z,s}^{1}$ &  $\sigma_{z,s}^{2}$ &  $\sigma_{z,s}^{3}$ \\
  \hline
  $a_1$ & -0.89 & 0.83 & 0.89 & 0.86 \\
 $\eta_1$ & 0.60 & -0.74 & -0.70 & -0.58 \\
 \hline
\end{tabular}
 \caption{Correlation coefficients between IA and photo-z model parameters displaying significant degeneracies under fiducial IA parameter set 4. Note that we list the correlation coefficient of $\eta_1$ and $\sigma_{z,s}^{3}$ although it is below our nominal |0.6| threshold for completeness.}
  \label{tab:corr_coeffs_a1eta1}
 \end{center}
\end{table}

The degeneracy between $a_1$ and the mean of tomographic source bins has been noted elsewhere in the literature (e.g., \cite{wright2020kids, stolzner2021self}). In \cite{fischbacher2023redshift} a degeneracy between $a_1$ and both the mean and width of source mean redshift bins was demonstrated in a cosmic shear only analysis, specifically for lower-redshift tomographic bins. We corroborate this finding, noting also that the strong correlations between $a_1$ and our redshift parameters are clustered towards lower $z$ bins. It is perhaps not surprising then that this effect appears to extend to degeneracies between $\eta_1$ and source redshift distribution parameters.

The degeneracy between $\eta_2$ and $\sigma_{z,s}^1$ is visualized in Figure~\ref{fig:degfind1}, where we show joint forecast constraints for $\eta_2$ and each of the $\sigma_{z,s}^i$ parameters within IA parameter set 1 (which has $a_2$, $\eta_2$, and $b_{\rm TA}$ non-zero). We see that whilst there is a strong negative degeneracy between $\eta_2$ and $\sigma_{z,s}^1$, this trend weakens and then reverses as we move to the higher $z$ source bins. To understand this, we should recall that parameterization of equation \ref{eq:A2} includes a pivot redshift, set here to $z_0=0.62$. The transition from negative to positive degeneracy between $\eta_2$ and $\sigma_{z,s}^i$ corresponds to the transition whereby the source bin $i$ transitions from $\bar{z}<0.62$ to $\bar{z}>0.62$ (see Figure~\ref{fig:dndzs}). For a source bin with $z<z_0$, reducing $\eta_2$ increases the overall amplitude of Equation~\eqref{eq:A2}; for $z>z_0$, the opposite effect occurs.

Finally, in the case of degeneracies of $b_{\rm TA}$ with $\Delta \bar{z}^i$ and with $\sigma_{z,s}^i$, we look to equation~\eqref{eq:bTA}. We see that $b_{\rm TA}$ enters our equations as a direct multiplier of $A_1(z)$, meaning that it will have strong negative degeneracy with $a_1$ when the fiducial value of $b_{\rm TA}$ is non-zero (and we directly verify that this is the case.) As a result of the degeneracy between $b_{\rm TA}$ and $a_1$, it follows that the primary strong degeneracies we observe for $a_1$ (with $\Delta \bar{z}_s^i$ and $\sigma_{z,s}^i$) will also be present with respect to $b_{\rm TA}$ at some level. Figure \ref{fig:degbTAzssigzs} shows the degeneracies between $b_{\rm TA}$ and relevant redshift parameters under IA parameter set 10 (all 5 TATT parameters non-zero, $a_1$ negative and the remainder positive).

We note briefly that although we agree with the cosmic shear only analysis of \cite{fischbacher2023redshift}, in comparing with that work we see the importance of considering different fiducial sets of IA parameters within the allowed parameter space, as these impact the significant degeneracies which are present. We reproduce their findings for the small number of fiducial parameter sets they consider (effectively corresponding to a redshift-independent TATT and an NLA model), and are able to further identify the above degeneracies involving $\eta_2$ and $b_{\rm TA}$ as a result of considering a broader set of fiducial IA parameters. 

\section{Biases under IA and photo-z mis-specification}
\label{sec:bias}
Now that we have identified three distinct scenarios where there is significant degeneracy between IA and photo-z parameters, we explore each of these scenarios. In each case, we consider the degeneracy in play within the scenario where the truth IA parameter set is that under which we identified the degeneracy in Section \ref{sec:finddegs}. We seek to determine then whether various mis-specifications of one or more of the highly correlated IA and photo-z parameters results in significant cosmological parameters bias, in the best-fit model being empirically identifiable as a poor fit to the data via the $\chi^2_{\rm DOF}$ metric, or both. The hypothetical worst case scenario is that in which the cosmological parameter bias is high, but $\chi^2_{\rm DOF}$ indicates an acceptable fit: this is the scenario where in a real analysis we could have significant cosmological parameter bias due to IA and photo-z mis-specification, but we would not empirically identify this issue via $\chi^2_{\rm DOF}$. 

The headline results of this section, which are discussed in much more detail below, are summarized in Table \ref{tab:summary_biases}. For each analysis scenario considered below, this table presents the following summary statistic:
\begin{equation}
    {\rm Max}(\Delta (p_1 - p_2))\times \left( \frac{|p_3^{\rm assumed} - p_3^{\rm true}|}{|p_3^{\rm true}|} \right)^{-1}
    \label{eq:summary_stat}
\end{equation}
where $p_1$ and $p_2$ are either $S_8$ and $\Omega_{\rm M}$ or $w_0$ and $w_a$ respectively, $\Delta (p_1 - p_2)$ is  given in fractions of $\sigma$, $p_3$ is the parameter being mis-specified in the given scenario, and the maximum is taken over mis-specifications which result in $\chi^2_{\rm DOF} \le 1.1$. In other words, equation \ref{eq:summary_stat} provides the maximum parameter bias normalized by the fractional  magnitude of the parameter mis-specification which causes that bias, for cases with acceptable goodness-of-fit (where such cases exist). As such, it attempts to summarize the sensitivity of parameter bias to the level of mis-specification, in scenarios which would not be empirically flagged as bad fits by $\chi^2_{\rm DOF}$, where a higher value means a greater sensitivity. Note that in the case where the true value of $p_3$ is 0 (for $\eta_1$ and $\eta_2$), we instead substitute unity in the above expression as a reasonable fiducial value of these parameters. 

We see immediately from Table \ref{tab:summary_biases} that the scenarios of greatest sensitivity which also have acceptable $\chi^2_{\rm DOF}$ values are those related to the degeneracy identified between $b_{\rm TA}$ and $\sigma_{z,s}^{2,3}$ when $a_1<0$. There is also a comparative sensitivity in the case where $\sigma^{1-3}_{z,s}$ is mis-specified under a correct IA modelling with NLA-z.

\begin{table*}
\begin{center}
\begin{tabular}{ |c|c|c|c|c|c|c|c|  }
 \hline
True IA & Assumed IA & True PZ & Assumed PZ & \makecell{$\Delta (S_8 - \Omega_{\rm M})$ \\ sensitivity statistic \\ (Equation \ref{eq:summary_stat})} & \makecell{$\Delta (w_0 - w_a)$ \\ sensitivity statistic \\ (Equation \ref{eq:summary_stat})}  & Figure & Section \\
 \hline
NLA-z  &  NLA-z  & \makecell{binned shift \\ $\sigma_{z,s}^{1-3}$ mis-spec}  & binned shift & 0.02 & 0.1 & \ref{fig:sigz1-3_NLA_misspec} & \ref{subsubsec:misspecsigz1-3}\\

\makecell{NLA-z \\ $\eta_1$ mis-spec} & NLA & binned shift & binned shift & 0.004 & 0.01 & \ref{fig:eta1_NLAz_misspec} & \ref{subsubsec:misspeceta1}\\
\hline
TATT ($a_1>0$) & TATT ($a_1>0$) & \makecell{binned shift \\ $\sigma_{z,s}^{1}$ mis-spec} & binned shift & - & - & \ref{fig:sigz1only_misspec} & \ref{subsubsec:mispsecsig1} \\

TATT ($a_1>0$) & NLA-z &\makecell{binned shift \\ 
$\sigma_{z,s}^{1}$ mis-spec} & binned shift & - & - & \ref{fig:sigz1_misspec_NLAz}& \ref{subsubsec:mispsecsig1} \\

\makecell{TATT ($a_1>0$) \\ $\eta_2$ mis-spec} & NLA-z  & binned shift  & binned shift  & - & - & \ref{fig:eta2_misspec_NLAz} & \ref{subsubsec:mispseceta2}  \\
\hline
TATT ($a_1<0$) & TATT ($a_1<0$) & \makecell{binned shift \\ $\sigma_{z,s}^{2,3}$ mis-spec} & binned shift & 0.02 & 0.1 & \ref{fig:sig23_misspec_TATT} & \ref{subsubsec:sig23misspec} \\

TATT ($a_1<0$) & \makecell{TATT ($a_1<0$)\\ Fix $b_{\rm TA}=1$} & \makecell{binned shift \\ $\sigma_{z,s}^{2,3}$ mis-spec} & binned shift & 0.3 & 0.07 & \ref{fig:sig23_misspec_TATT_bTAfixed} & \ref{subsubsec:sig23misspec} \\

\makecell{TATT ($a_1<0$) \\ $b_{\rm TA}=1$ mis-spec} & \makecell{TATT ($a_1<0$) \\ Fix $b_{\rm TA}=1$}  & binned shift & binned shift & 0.08 & 0.06 & \ref{fig:bTA_misspec_TATT} & \ref{subsubsec:bTAmisspec} \\
 \hline
\end{tabular}
 \caption{Summary of the results of Section \ref{sec:bias}. The fifth and sixth columns provide the values of the summary statistic defined in equation \ref{eq:summary_stat}, which is designed to quantify the sensitivity of the cosmological parameters to the IA or photo-z model mis-specification in each scenario investigated. A higher value indicates higher sensitivity to this mis-specification scenario. No value indicates that there were no scenarios considered which resulted in an acceptable $\chi^2_{\rm DOF}$ value. See discussion at the beginning of Section \ref{sec:bias} for more detail.}
  \label{tab:summary_biases}
 \end{center}
\end{table*}
We now proceed to discuss these scenarios.

\subsection{Mis-specifying $\sigma_{z,s}^i$ and $\eta_1$ under truth NLA-z IA model}
\label{subsec:sigz1eta1}

We first consider the degeneracy between the mean and variance of the source galaxy redshift distributions (particularly for lower-$z$ bins) and the IA amplitude and redshift dependence, in the case in which the IA is well described by a redshift-dependent NLA model only (i.e. true values of $a_2$ and $b_{\rm TA}$ are zero).

\subsubsection{Mis-specified $\sigma_{z,s}^{i=1-3}$}
\label{subsubsec:misspecsigz1-3}
\begin{figure*}
\centering
\subfigure{\includegraphics[width=0.45\textwidth]{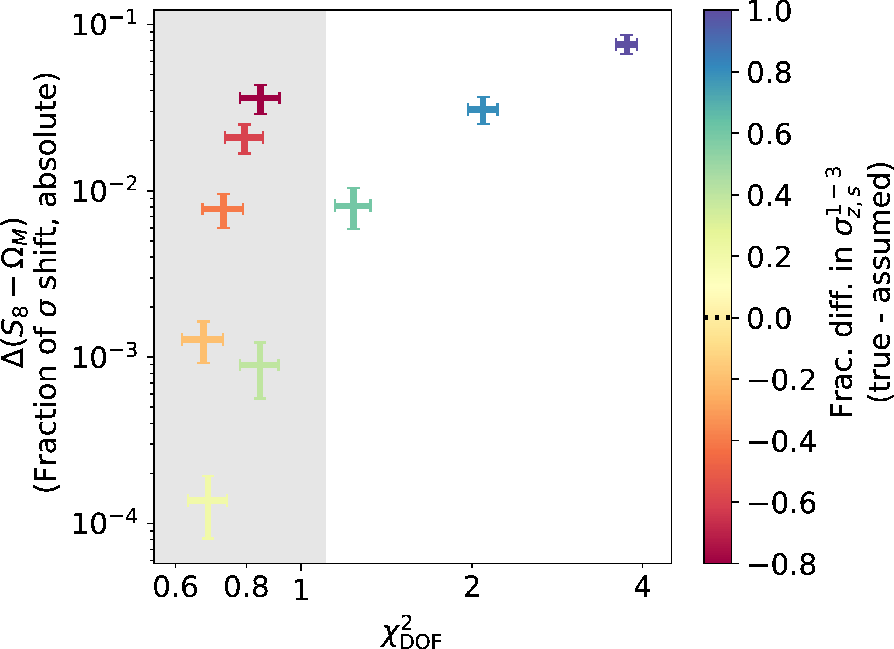}}
\subfigure{\includegraphics[width=0.45\textwidth]{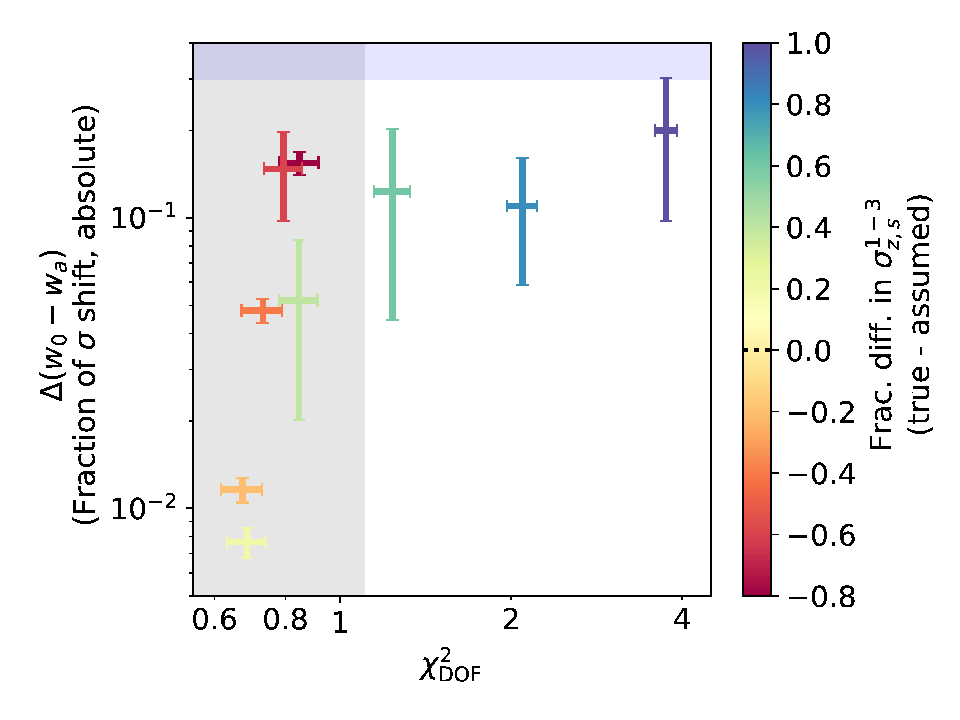}}
\caption{Effect of mis-specifying $\sigma_{z,s}^{1-3}$, with true and assumed IA scenario described by redshift-dependent NLA (IA parameter set 4). The assumed value is $\sigma_{z,s}^{1-3} = 0.05$. The grey shaded region indicates $\chi^2_{\rm DOF} \le 1.1$. The blue shaded region indicates parameter bias of at least 0.3$\sigma$. See Section \ref{subsubsec:misspecsigz1-3} for more details.}
\label{fig:sigz1-3_NLA_misspec}
\end{figure*}

In the first instance, we consider the effect of mis-specifying the photo-z parameters $\sigma_{z,s}^{i={1-3}}$. Specifically, we take:

\begin{center}
\begin{tabular}{|c|c|c|}
 \hline
 & Truth & Assumed \\
 \hline
IA  & \makecell{NLA-z\\(par set 4)}  & \makecell{NLA-z\\(par set 4)} \\
 \hline
Photo-z & \makecell{binned shift \\ (baseline pars) \\ $\sigma_{z,s}^{1-3}$ mis-spec} & \makecell{binned shift \\ (baseline pars)} \\
\hline
\end{tabular}
\end{center}
Figure \ref{fig:sigz1-3_NLA_misspec} shows the results of this mis-specification. As this is the first of numerous figures with similar format, we pause to indicate several conventions we use here and in similar figures below. First, recall as stated above that here and in all similar figures below, errors on $\Delta (S_8 - \Omega_{\rm M})$ and $\Delta (w_0 - w_a)$ account for numerical differentiation error whilst errors on $\chi^2_{\rm DOF}$ describe the variation in this metric as a result of a noisy data vector. We will indicate the level of parameter mis-specification in terms of fractional difference between true and assumed value:
\begin{equation}
    \frac{p_{\rm true} - p_{\rm assumed}}{p_{\rm assumed}}.
    \label{eq:frac_diff}
\end{equation}
The exception will be when $p_{\rm assumed}=0$ (not the case here), where we will instead show dependence on $p_{\rm true}$ itself. We indicate $\chi^2_{\rm DOF} \le 1.1$ (our subjective metric for acceptable goodness-of-fit) via grey shading, and we indicate a 2D parameter bias exceeding $0.3\sigma$ via blue shading. A mis-specification which results in a parameter bias and goodness-of-fit in the overlap of these regions is the scenario of greatest concern.

We do not see any data points in this concerning overlap region in Figure \ref{fig:sigz1-3_NLA_misspec}. Neither the bias in $S_8 - \Omega_{\rm M}$ nor $w_0 - w_a$ is severe, with shifts in both below the $0.3\sigma$ level. $\chi^2_{\rm DOF}$ values indicate a good fit across most mis-specifications considered, although in the case where the underlying true $\sigma_{z,s}^{i=1-3}$ are increasingly greater than their assumed values, we do begin to see $\chi^2_{\rm DOF}$ values indicative of a bad fit.

We see that there are two `branches' in each plane of Figure \ref{fig:sigz1-3_NLA_misspec}: one representing the case where the true $\sigma_{z,s}^{1-3}$ are overestimated and one where they are underestimated in the assumed model. The case of overestimating the photo-z variances produces better fits in terms of $\chi^2_{\rm DOF}$ than in the underestimation case, although $\Delta (S_8 - \Omega_{\rm M})$ and $\Delta(w_0 - w_a)$ values are comparable between branches. The effect of symmetric mis-specifications of source variance parameters about their correct value would be expected the same for Gaussian-like source sample redshift distributions, where we would expect the central moments of the source sample redshift distributions to be even powers of the standard deviation. However, the asymmetric integration limits in the lensing kernel (equation \ref{Wdef}) will induce an asymmetric dependence of the cosmological parameter bias on (under/over)estimated standard deviations even in the linearized Fisher bias formalism (equation \ref{biasF}). This general pattern of asymmetric branching about the instance where true and assume parameter values coincide is something we will see repeatedly. 

\begin{figure*}
\centering
\subfigure{\includegraphics[width=0.3\textwidth]{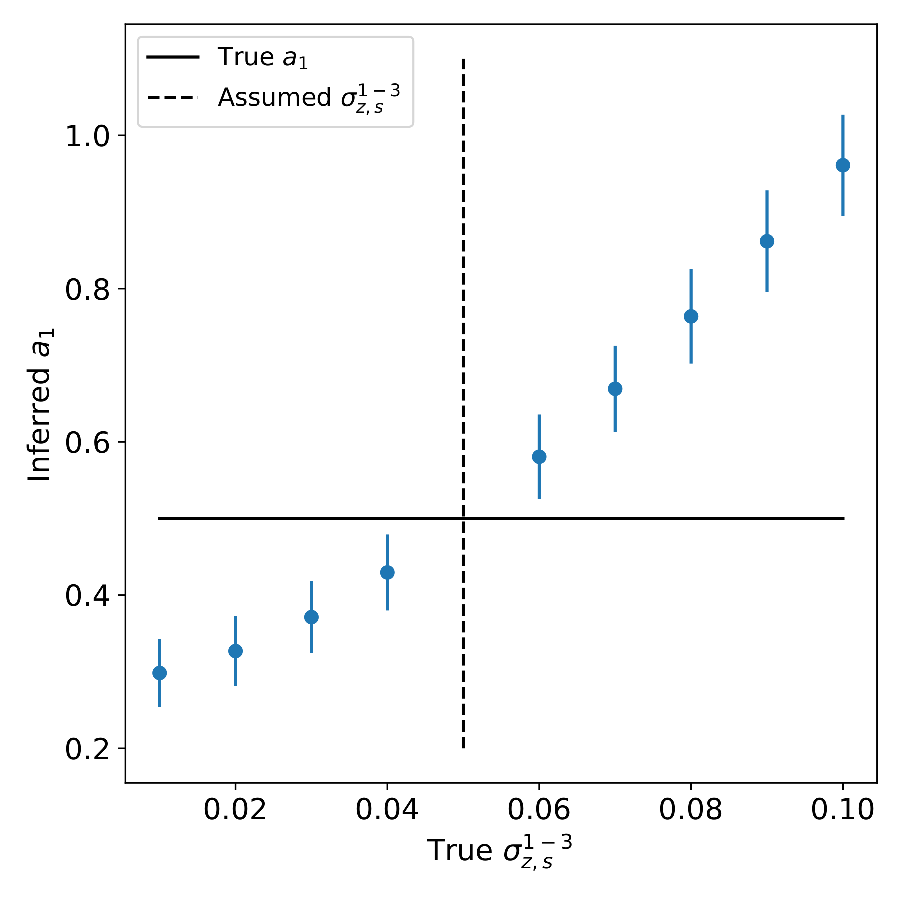}}
\subfigure{\includegraphics[width=0.3\textwidth]{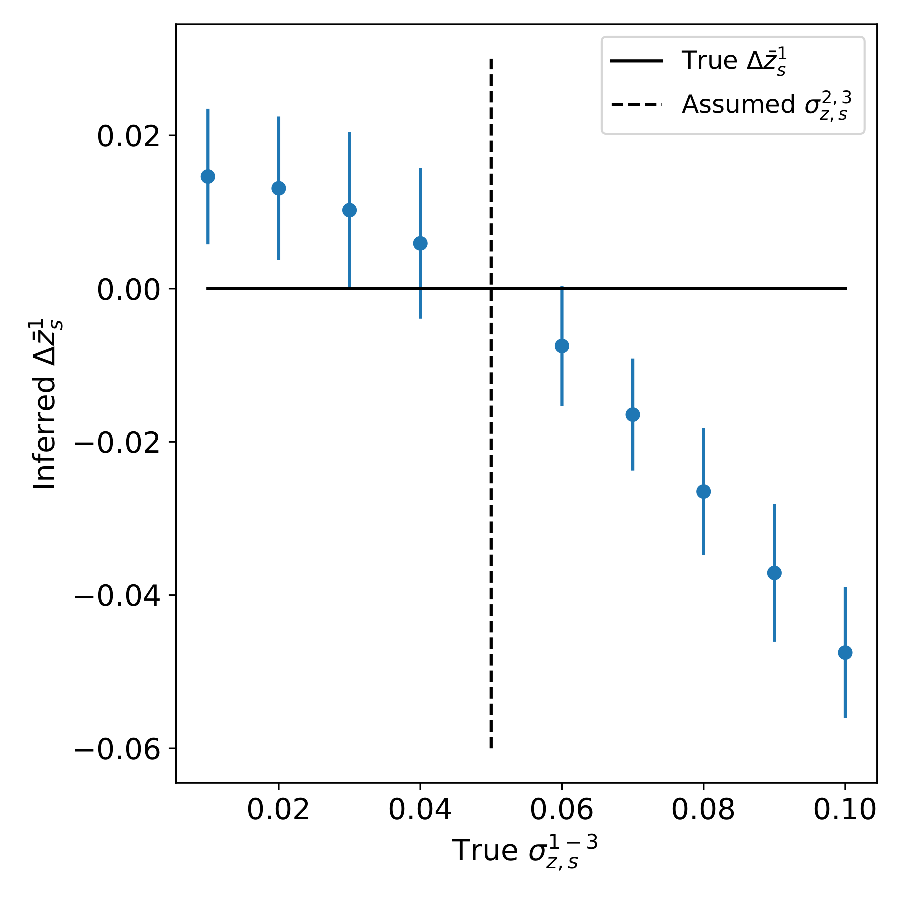}}
\subfigure{\includegraphics[width=0.3\textwidth]{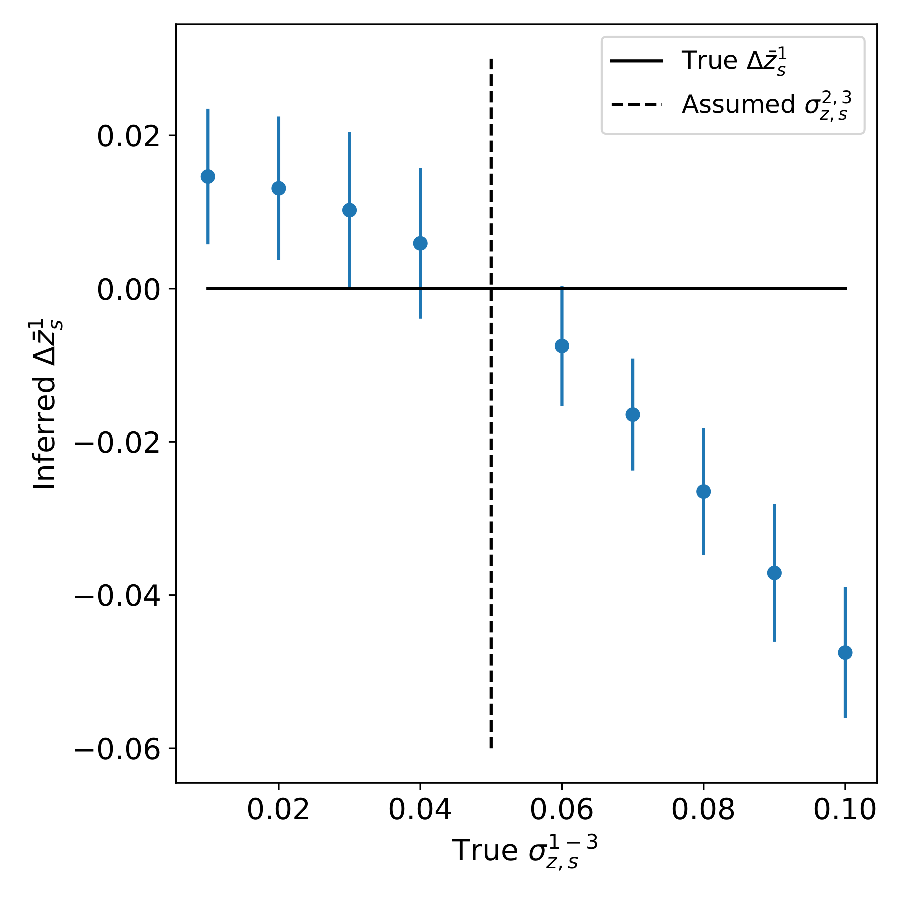}}
\caption{Bias to a subset of non-cosmological parameters due to mis-specifying $\sigma_{z,s}^{1-3}$, with true and assumed IA scenario described by redshift-dependent NLA (IA parameter set 4). We see that although cosmological parameter bias levels are acceptable (see Figure \ref{fig:sigz1-3_NLA_misspec}), IA and photo-z parameters are severely biased. Error bars represent posterior forecast uncertainties rather than Fisher instability uncertainties in this case. See Section \ref{subsubsec:misspecsigz1-3} for more details.}
\label{fig:IA_par_biases}
\end{figure*}

In examining the broader set of parameter biases, we find significant biases in the inferred values of $a_1$, $\eta_1$, and $\Delta \bar{z}_s^1$, with all three parameters being inferred to be several $\sigma$ away from their true values at the extreme end of the $\sigma_{z,s}^{1-3}$ mis-specifications considered. This is illustrated in Figure \ref{fig:IA_par_biases}. Even though the cosmological parameters are not significantly biased in this scenario, it would thus be dangerous to interpret the inferred values of the IA and photo-z parameter values themselves as being physically meaningful and transferable to other analyses e.g. as a means of informing priors.

\subsubsection{Mis-specified $\eta_1$}
\label{subsubsec:misspeceta1}
Under the same IA truth set-up, we consider now the case where the redshift distribution is correctly specified but a redshift-independent NLA model is incorrectly assumed. The degeneracies between $\eta_1$ and $\bar{z}_s^i$ may induce bias in the latter, which could propagate into the cosmological parameters. Specifically, we look at:
\begin{center}
\begin{tabular}{|c|c|c|}
 \hline
 & Truth & Assumed \\
 \hline
IA  & \makecell{NLA-z\\(par set 4) \\$\eta_1$ mis-spec}  & \makecell{NLA\\(par set 5)} \\
 \hline
Photo-z & \makecell{binned shift \\ (baseline pars)} & \makecell{binned shift \\ (baseline pars)} \\
\hline
\end{tabular}
\end{center}

\begin{figure*}
\centering
\subfigure{\includegraphics[width=0.45\textwidth]{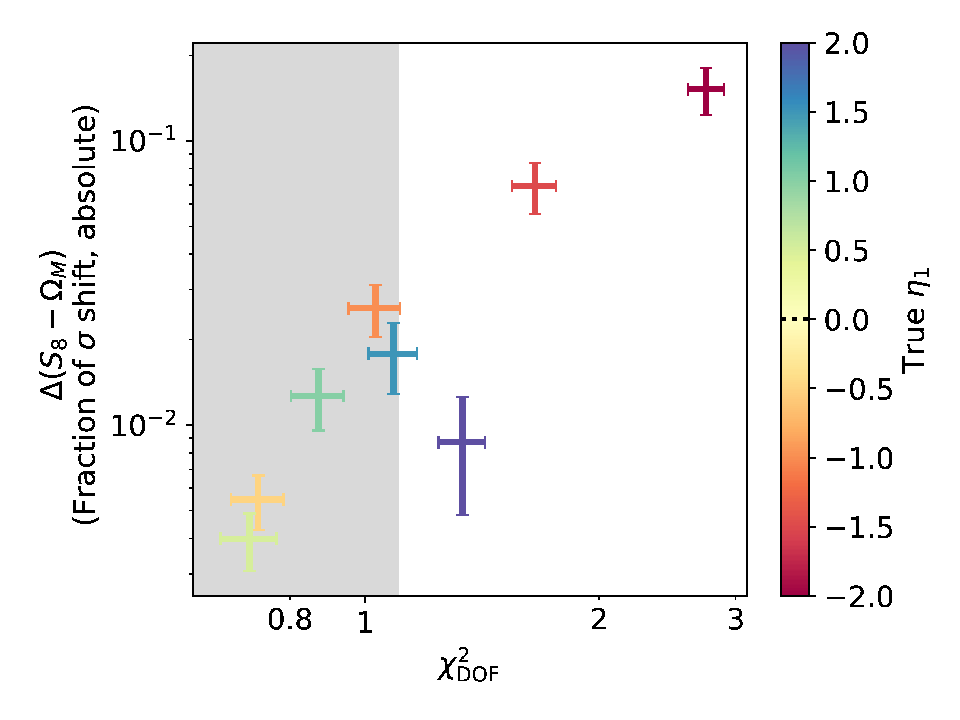}}
\subfigure{\includegraphics[width=0.45\textwidth]{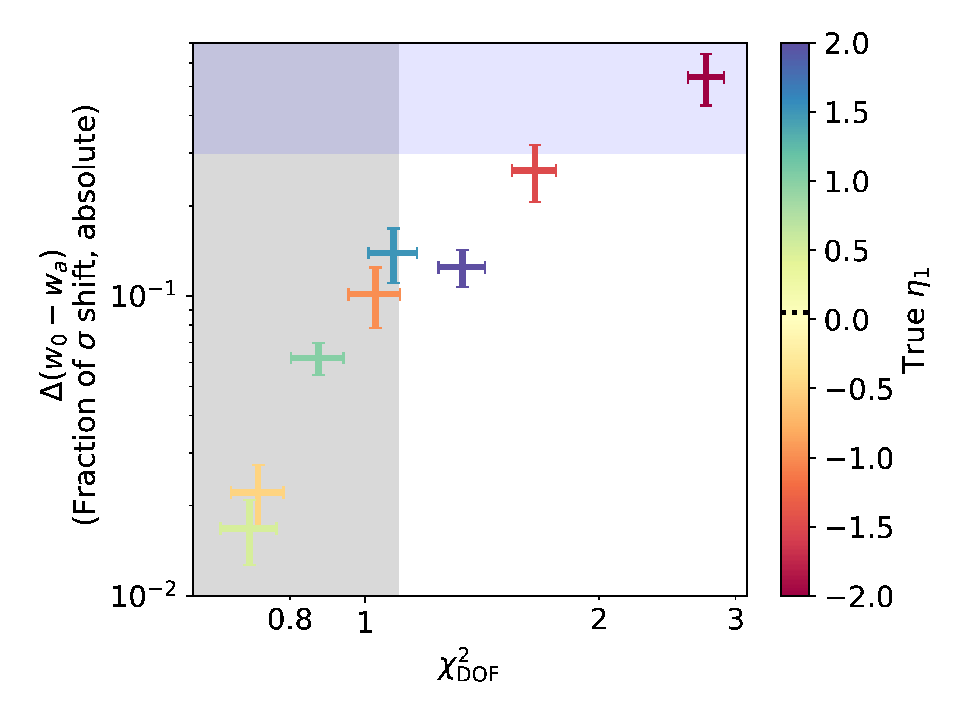}}
\caption{Inferred cosmological parameters biases and goodness-of-fit of inferred parameters in the scenario where the true IA is given by a redshift-dependent NLA model but the analysis assumes a redshift-independent NLA model with $\eta_1=0$ (indicated on the color-bar by a dashed black line). The grey shaded region indicates $\chi^2_{\rm DOF} \le 1.1$. The blue shaded region indicates parameter bias of at least 0.3$\sigma$. See Section \ref{subsubsec:misspeceta1} for more details. }
\label{fig:eta1_NLAz_misspec}
\end{figure*}

Result are shown in Figure \ref{fig:eta1_NLAz_misspec}. We see that once again we have two branches of behavior: that where $\eta_1^{\rm truth}<\eta_1^{\rm assumed}$ (in this case meaning $\eta_1^{\rm truth} < 0$) and vice versa. When true $\eta_1$ is negative, $\Delta (S_8 - \Omega_{\rm M})$ and $\Delta (w_0 - w_a)$ rise steadily with increasing true $|\eta_1|$, exceeding the $0.3 \sigma$ threshold in the $\Delta (w_0 - w_a)$ plane when $\eta_1=-2.0$. The case of positive $\eta_1$ (IA amplitude increasing with redshift) is more complex, with non-monotonic behavior in $\Delta (S_8 - \Omega_{\rm M})$ (and a corresponding, but less dramatic, feature in $\Delta (w_0 - w_a)$). 

These results suggest that if the true IA amplitude is decreasing sharply with redshift but we assume it to be redshift-independent, the result may be significant bias to inferred cosmological parameters, particularly $w_0$ and $w_a$. However, when $\Delta (w_0 - w_a)$ exceeds $0.3 \sigma$, we note that $\chi^2_{\rm DOF}$ is above our threshold acceptable value of 1.1 and therefore in principle this mis-specification should be identifiable. As an aside, we note that this is an example in which intuition around $\chi^2_{\rm DOF}$ values built from scenarios with smaller numbers of degrees of freedom (where, e.g.  $\chi^2_{\rm DOF}$ approaching 2 may correspond to acceptable p-values) would fail to identify an issue. This reinforces the need to take proper account of the dependence of the behavior of the $\chi^2$ distributions on $n_{\rm DOF}$ in modern cosmological analysis. 

Biases are again present in this scenario in the inferred mean shifts of the source tomographic redshift bins, most severely $\Delta \bar{z}_s^1$, as well as in $a_1$. However, compared to the case of mis-specifying $\sigma_{z,s}^{i=1-3}$ above, the absolute degree of this biasing is noticeably lower. We conjecture that the relative lack of flexibility in the redshift-dependent systematics parameterization (notably the fixing of $\eta_1$ in the assumed model) results in the effect of mis-specification being pushed into the cosmological sector rather than being absorbed by the photo-z and IA parameters (as in the above case where $\eta_1$ is included in the assumed model and marginalized over). These results caution against the use of a redshift-independent NLA model in Stage-IV surveys. Although redshift-independent NLA is by no means a universal choice in Stage III analyses, some recent analyses of Stage III cosmic shear do incorporate it as a modelling option (see e.g. \citep{li2023kids, secco2022dark}). It is worth bearing in mind that this is unlikely to be an appropriate choice going forward to Stage IV unless strong priors can be places on the redshift-dependence of tidal alignment {\it a priori}.

\subsection{Mis-specifying  $\sigma_{z,s}^1$ and $\eta_2$ under IA truth model TATT}
\label{subsec:eta2}

We now move away from the case in which the truth IA model is well-described by redshift-dependent NLA, and consider cases in which the true IA behavior includes tidal torquing and source density weighting. Within this, we are first interested in understanding the effect of the degeneracy identified in Section \ref{sec:finddegs} between $\eta_2$ and $\sigma_{z,s}^1$. 

\subsubsection{Mis-specification of $\sigma_{z,s}^1$}
\label{subsubsec:mispsecsig1}
We consider initially a scenario where $\sigma_{z,s}^1$ is fixed and mis-specified, whereas all parameters of the TATT model including $\eta_2$ are allowed to vary. Specifically, we take:

\begin{center}
\begin{tabular}{|c|c|c|}
 \hline
 & Truth & Assumed \\
 \hline
IA  & \makecell{TATT\\(par set 1)}  & \makecell{TATT\\(par set 1)} \\
 \hline
Photo-z & \makecell{binned shift \\ (baseline pars) \\ $\sigma_{z,s}^1$ mis-spec} & \makecell{binned shift \\ (baseline pars)} \\
\hline
\end{tabular}
\end{center}

In this set up, we expect a possible cosmological parameter bias as a result of the fact that mis-specifying $\sigma_{z,s}^1$ would be expected to bias $\eta_2$ due to their tight correlation, with biases to $\eta_2$ then potentially leading to biases in cosmological parameters.

\begin{figure*}
\centering
\subfigure{\includegraphics[width=0.45\textwidth]{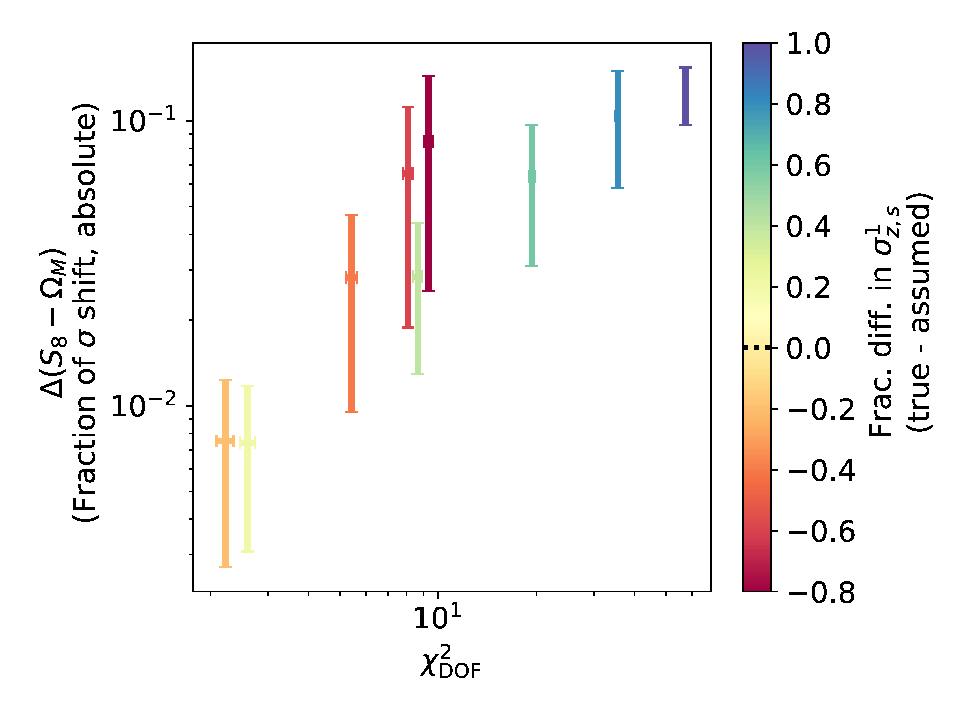}}
\subfigure{\includegraphics[width=0.45\textwidth]{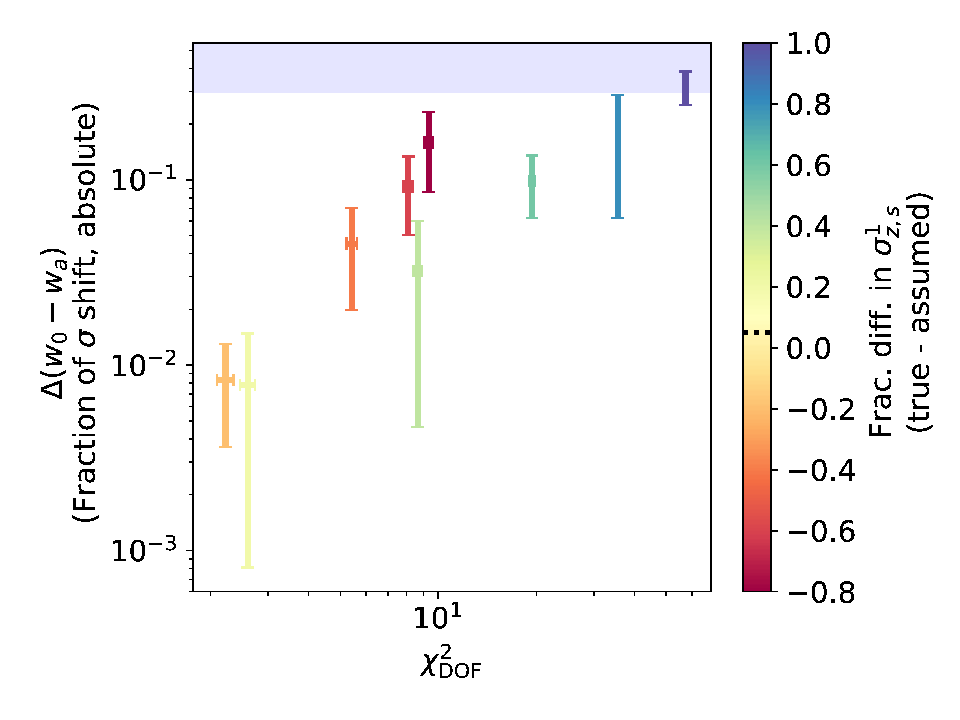}}
\caption{Effect of mis-specification of $\sigma_{z,s}^1$, with true and assumed IA model given by TATT with all parameters positive non-zero (IA parameter set 1) and $\sigma_{z,s}^{1, \rm{assumed}} = 0.05$. The blue shaded region indicates parameter bias of at least 0.3$\sigma$. The $\chi^2_{\rm DOF}$ is greater than our threshold of $ 1.1$ for all points, hence no grey shading is included. See Section \ref{subsubsec:mispsecsig1} for more details.}
\label{fig:sigz1only_misspec}
\end{figure*}

Results can be seen in Figure \ref{fig:sigz1only_misspec}. We see that once again we have two `branches', corresponding to $\sigma_{z,s}^{1, {\rm true}}<\sigma_{z,s}^{1, {\rm assumed}}$ and $\sigma_{z,s}^{1, {\rm true}}>\sigma_{z,s}^{1, {\rm assumed}}$. For the range of true $\sigma_{z,s}^1$ values considered here, the $\chi^2_{\rm DOF}$ metric responds much more strongly to the mis-specification than do the cosmological parameter bias metrics. For none of the $\sigma_{z,s}^1$ truth values considered does $\Delta (S_8 - \Omega_{\rm M})$ approach the $0.3 \sigma$ bias level, and only for the most extreme cases does $\Delta (w_0 - w_a)$. However, all but the smallest absolute differences in true and assumed $\sigma_{z,s}^1$ produce $\chi^2_{\rm DOF}$ which are indicative of the inferred parameter values being a bad fit to the the truth datavector. 

To understand why this occurs, we once again consider the non-cosmological parameter biases which are induced by this mis-specification. We find that, unsurprisingly, $\eta_2$ is subject to the most severe biases. However, it is evident from Figure \ref{fig:sigz1only_misspec} that this does not then propagate into biases in the cosmological parameters of interest. In examining the data vector corresponding to the inferred parameters, we see that there is a very strong impact on those cosmic shear spectra which include source bin 1, with a much more moderate impact on all other spectra. We conclude that this impact on the data vector is simply not degenerate with the effect of varying our cosmological parameters of interest. However, the modelled data vector is significantly discrepant from the true data vector, resulting in a very high $\chi^2_{\rm DOF}$. This case is thus one in which the use of a $\chi^2_{\rm DOF}$ diagnostic would signal a problem when in fact cosmological parameter bias levels may be well within tolerance. Whether this is a positive or negative feature depends on the objective of the analysis at hand.

Would the impact of the mis-specification of $\sigma_{z,s}^1$ on the inferred cosmological parameter values be more severe if $\eta_2$ was fixed and hence unable to `absorb' the effect of the mis-specification? We investigate this possibility by considering an adjusted set-up where the assumed IA model is redshift-dependent NLA with fiducial parameter set 4 from Table \ref{tab:IA_model_vary}:
 
\begin{center}
\begin{tabular}{|c|c|c|}
 \hline
 & Truth & Assumed \\
 \hline
IA  & \makecell{TATT\\(par set 1)}  & \makecell{NLA-z\\(par set 4)} \\
 \hline
Photo-z & \makecell{binned shift \\ (baseline pars) \\ $\sigma_{z,s}^1$ mis-spec} & \makecell{binned shift \\ (baseline pars)} \\
\hline
\end{tabular}
\end{center}

This means that in the assumed IA model, $a_1$ and $\eta_1$ have the same fiducial values as in the truth IA model, but $a_2$, $\eta_2$, $b_{\rm TA}$ are erroneously set to 0. Note that this implies that in addition to the mis-specification of $\sigma_{z,s}^1$, there is a persistent mis-specification of $a_2$, $\eta_2$, and $b_{\rm TA}$. Although this makes interpretation of the effect of the $\sigma_{z,s}^1$ mis-specification more difficult, we consider this to be the most realistic scenario in which $\eta_2$ would be fixed in an analysis.

\begin{figure*}
\centering
\subfigure{\includegraphics[width=0.45\textwidth]{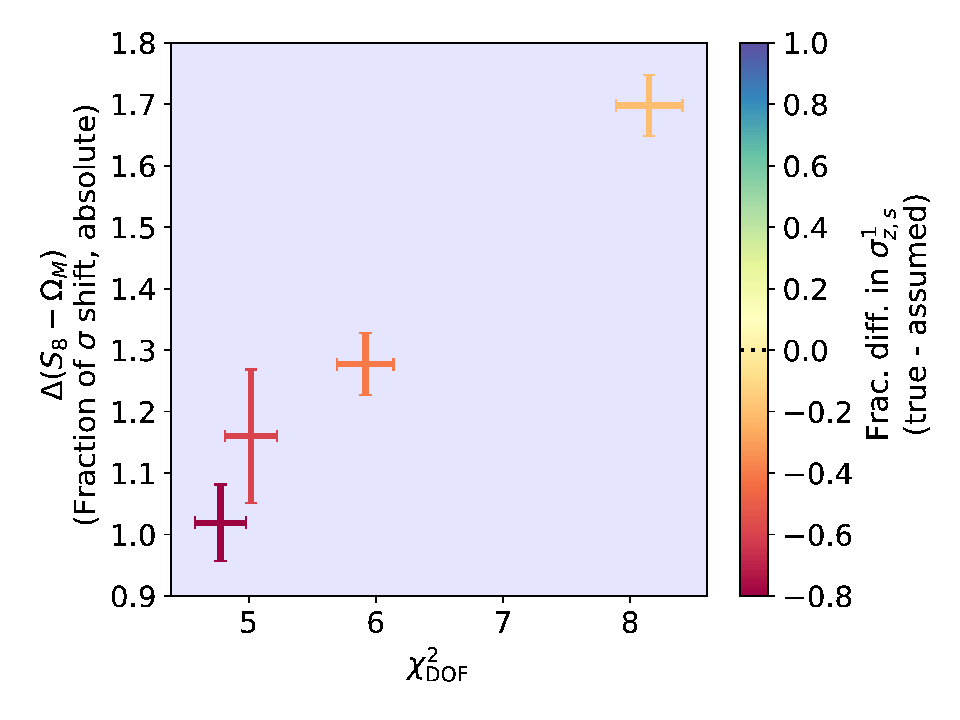}}
\subfigure{\includegraphics[width=0.45\textwidth]{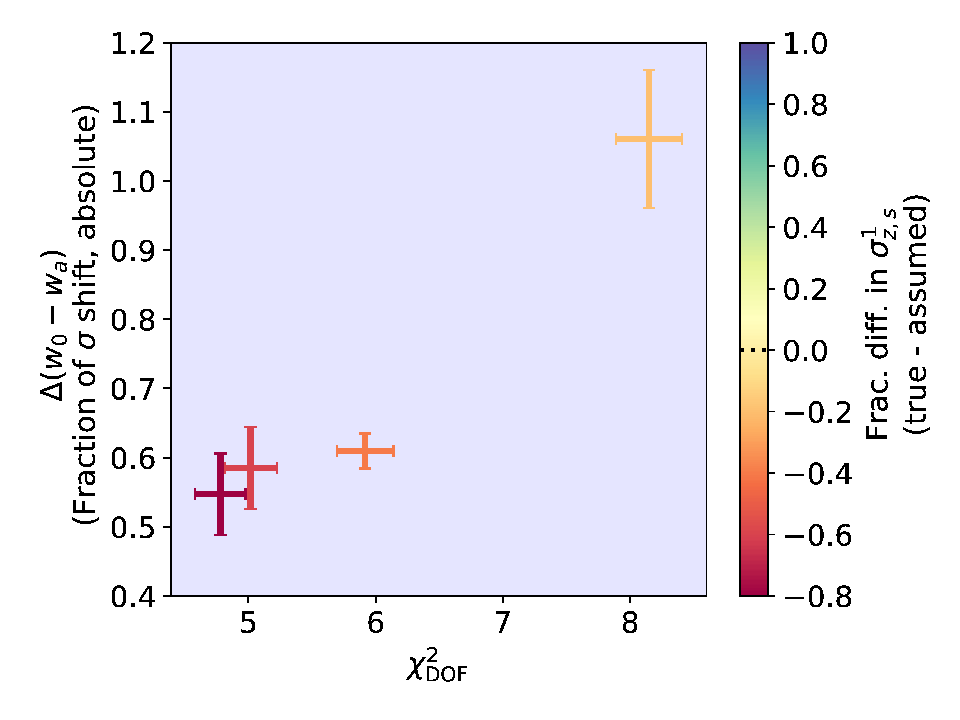}}
\caption{Effect of mis-specification of $\sigma_{z,s}^1$, with true IA model given by TATT with all parameters positive non-zero (IA parameter set 1), assumed IA model given by NLA-z with the same $a_1$ and $\eta_1$ values as in the true TATT model (IA parameter set 4), and $\sigma_{z,s}^{1, \rm{assumed}} = 0.05$.  The blue shading indicates parameter bias of at least 0.3$\sigma$ (true for all points in this scenario). $\chi^2_{\rm DOF}$ is greater than our threshold of $ 1.1$ for all points, hence no grey shading is included. See Section \ref{subsubsec:mispsecsig1} for more details.  Note that we maintain the same color bar scaling as Figure \ref{fig:sigz1only_misspec} for easy comparison. }
\label{fig:sigz1_misspec_NLAz}
\end{figure*}

The result is shown in Figure \ref{fig:sigz1_misspec_NLAz}. We see that the values of $\Delta (S_8 - \Omega_{\rm M})$ and $\Delta (w_0 - w_a)$ are much more severe than in the previous case, while values of $\chi^2_{\rm DOF}$ are of the same order of magnitude. Interestingly, for both $\Delta (S_8 - \Omega_{\rm M})$ and $\Delta (w_0 - w_a)$, the smallest parameter bias is given when $\sigma_{z,s}^{1, {\rm true}}=0.01$, despite the fact that this is the case where $\sigma_{z,s}^{1, {\rm assumed}} - \sigma_{z,s}^{1, {\rm true}}$ is greatest. The parameter biases in both planes (along with $\chi^2_{\rm DOF}$) actually increases as $\sigma_{z,s}^{1, {\rm true}}$ approaches $\sigma_{z,s}^{1, {\rm assumed}}$. We in fact only display results for a limited range of true $\sigma_{z,s}^1$ values here because when $\sigma_{\rm z,s}^{1, {\rm true}} \geq 0.05$, the corresponding biased value of $w_a$ so severe that we have a positive equation of state for dark energy. We infer that in the case where $\sigma_{z,s}^{1, {\rm assumed}}$ is greater than $\sigma_{z,s}^{1, {\rm true}}$, this somehow acts to partially compensate our incorrect assumption that $a_2$, $\eta_2$ and $b_{\rm TA}$ are zero.

Our conclusion from this exercise is that while there is some indication of interplay between the photo-z and IA mis-specification in this scenario, the bias in this set-up is predominantly sourced by the IA mis-specification. This is itself of interest considering the relatively moderate values of $a_2$, $\eta_2$ and $b_{\rm TA}$ assumed in the truth data vector (within the permitted values of current Stage III survey inference, see e.g. \cite{secco2022dark}), indicating that assuming a redshift-dependent NLA model will most likely be a highly inappropriate analysis choice for an early Stage IV 3$\times$2pt analysis.

\subsubsection{Mis-specification of $\eta_2$}
\label{subsubsec:mispseceta2}
We can also consider a scenario where it is $\eta_2$ which is mis-specified and $\sigma_{z,s}^1$ is not:

\begin{center}
\begin{tabular}{|c|c|c|}
 \hline
 & Truth & Assumed \\
 \hline
IA  & \makecell{TATT  \\ (par set 1) \\ $\eta_2$ mis-spec }  & \makecell{NLA-z \\ (par set 4)} \\
 \hline
Photo-z & \makecell{binned shift \\ (baseline pars)} & \makecell{binned shift \\ (baseline pars)} \\
\hline
\end{tabular}
\end{center}

In this case, the mis-specification of $\eta_2$ would be expected to propagate into the photometric redshift parameter sector due to the degeneracy with $\sigma_{z,s}^1$. While $\sigma_{z,s}^1$ is not itself permitted to vary, generic degeneracies between the $\sigma_{z,s}^i$ and the $\bar{z}_{z,s}^i$ parameters may result in a propagation of bias to the latter and hence potentially to the cosmological parameters of interest.

Once again we are in the situation where we are considering a mis-specification of one parameter ($\eta_2$), but, in order to consider a realistic analysis situation, we also must include a constant mis-specification of other parameters ($a_2$ and $b_{\rm TA}$). Thus even when $\eta_2^{\rm true}=\eta_2^{\rm assumed}=0$, we still expect biases. Our interest is in how the level of mis-specification of $\eta_2$ impacts the degree of these biases.

\begin{figure*}
\centering
\subfigure{\includegraphics[width=0.45\textwidth]{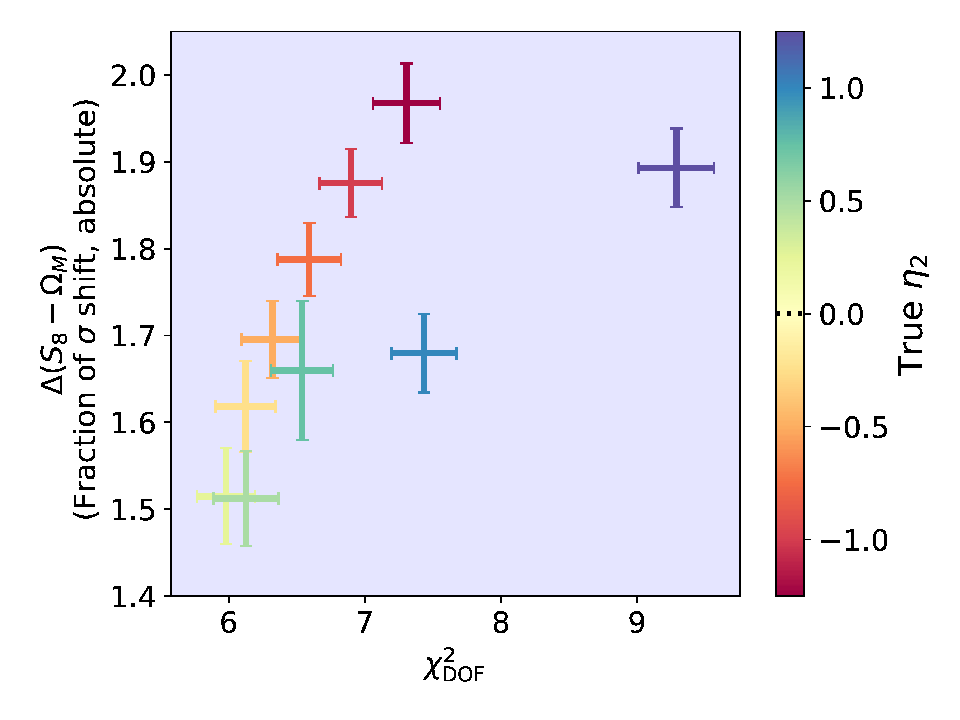}}
\subfigure{\includegraphics[width=0.45\textwidth]{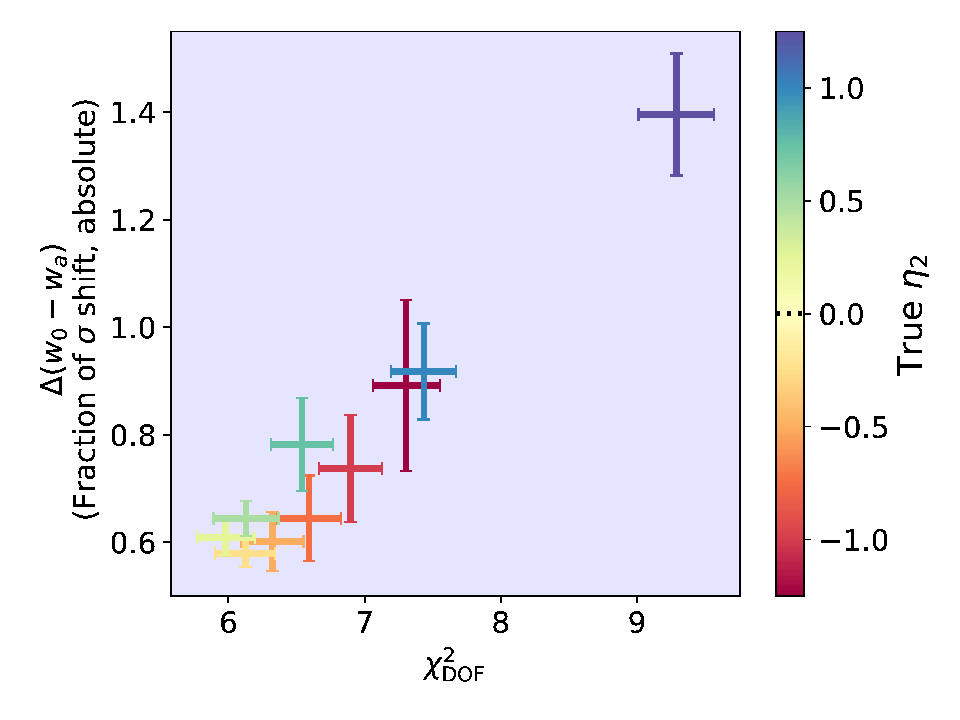}}
\caption{Effect of the mis-specification of $\eta_2$ under a truth IA model of TATT (IA parameter set 1) and an assumed IA model of redshift-dependent NLA (with $a_2$ and $b_{\rm TA}$ assumed to be 0, along with $\eta_2$ as indicated by the dashed line on the color-bar).  The blue shading indicates parameter bias of at least 0.3$\sigma$ (true for all points in this scenario). $\chi^2_{\rm DOF}$ is greater than our threshold of $ 1.1$ for all points, hence no grey shading is included. For further details, see Section \ref{subsubsec:mispseceta2}.}
\label{fig:eta2_misspec_NLAz}
\end{figure*}

Results are shown in Figure \ref{fig:eta2_misspec_NLAz}. We see that again, in all cases the overall levels of bias in $\Delta (S_8 - \Omega_{\rm M})$ and $\Delta (w_0 - w_a)$ are high, far exceeding the $0.3\sigma$ threshold in both 2D parameter spaces even in the situation where $\eta_2$ is actually correctly specified, reflective of the bias induced by the mis-specification of $a_2$ and $b_{\rm TA}$. Fortunately, this severe cosmological parameter bias corresponds to $\chi^2_{\rm DOF}$ values which are well above our goodness-of-fit threshold and the modelling mis-specification would likely be empirically identifiable in this case. 

If we attempt to discern the isolated effect of $\eta_2$, we see that the cosmological parameter bias level is quite sensitive to the level of $\eta_2^{\rm true}$ mis-specification, with the bias varying by $\approx 0.5\sigma$ over the mis-specification levels considered here. Although in this set-up the actual level of bias attributable to $\eta_2$ mis-specification (and the empirical identifiability of this mis-specification) is masked by the mis-specification of other IA parameters, the  mis-specification of $\eta_2$ could have a dominant and highly relevant impact in a scenario with smaller but non-zero truth value of $a_2$ and / or $b_{\rm TA}$. 

Looking at the biases to the non-cosmological parameters, we see that $a_1$ is significantly biased in such a way that does not scale significantly with $\eta_2^{\rm true}$, suggesting its bias is largely due to the constant bias in $a_2$ and $b_{\rm TA}$. The inferred values of $\eta_1$, $\Delta \bar{z}_s^1$ and $\Delta \bar{z}_s^5$ (the most biased of the mean shift parameters), on the other hand, are significantly dependent on $\eta_2^{\rm true}$.

\subsection{Mis-specifying $b_{\rm TA}$ and $\sigma_{z,s}^{2-3}$ under IA truth model TATT}
\label{subsec:bTA}

We look now at the last case of significant degeneracy identified in Section \ref{sec:finddegs}: that of $b_{\rm TA}$ with a subset of our photometric redshift parameters, most egregiously with $\bar{z}_s^1$, $\sigma_{s}^2$, and $\sigma_{s}^3$. It is worth stating for clarity that the base truth models for IA which we consider in this section have an important qualitative difference from those considered above, in that $a_1$ is negative whereas above it was positive. Although negative $a_1$ is perhaps not physically intuitive, it is an available region of IA parameter space from current Stage III results (see e.g. \cite{secco2022dark}) and so we consider it here.

\subsubsection{Mis-specification of $\sigma_{z,s}^{2-3}$}
\label{subsubsec:sig23misspec}
We first investigate the effect of mis-specification in $\sigma_{z,s}^{2-3}$, with the idea that the degeneracy of these parameters with $b_{\rm TA}$ may lead to bias in the latter and therefore also bias in other parts of parameter space. Specifically, we consider:

\begin{center}
\begin{tabular}{|c|c|c|}
 \hline
 & Truth & Assumed \\
 \hline
IA  & \makecell{TATT  \\ (par set 10)}  & \makecell{TATT  \\ (par set 10)} \\
 \hline
Photo-z & \makecell{binned shift \\ (baseline pars) \\ $\sigma_{z,s}^{2-3}$ mis-spec} & \makecell{binned shift \\ (baseline pars)} \\
\hline
\end{tabular}
\end{center}

\begin{figure*}
\centering
\subfigure{\includegraphics[width=0.45\textwidth]{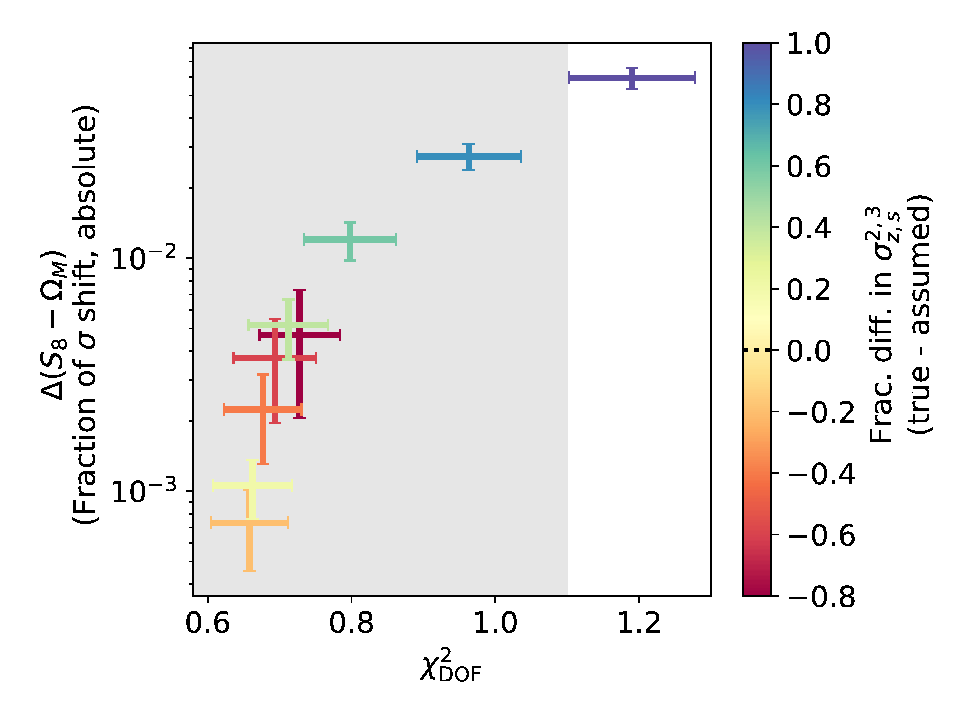}}
\subfigure{\includegraphics[width=0.45\textwidth]{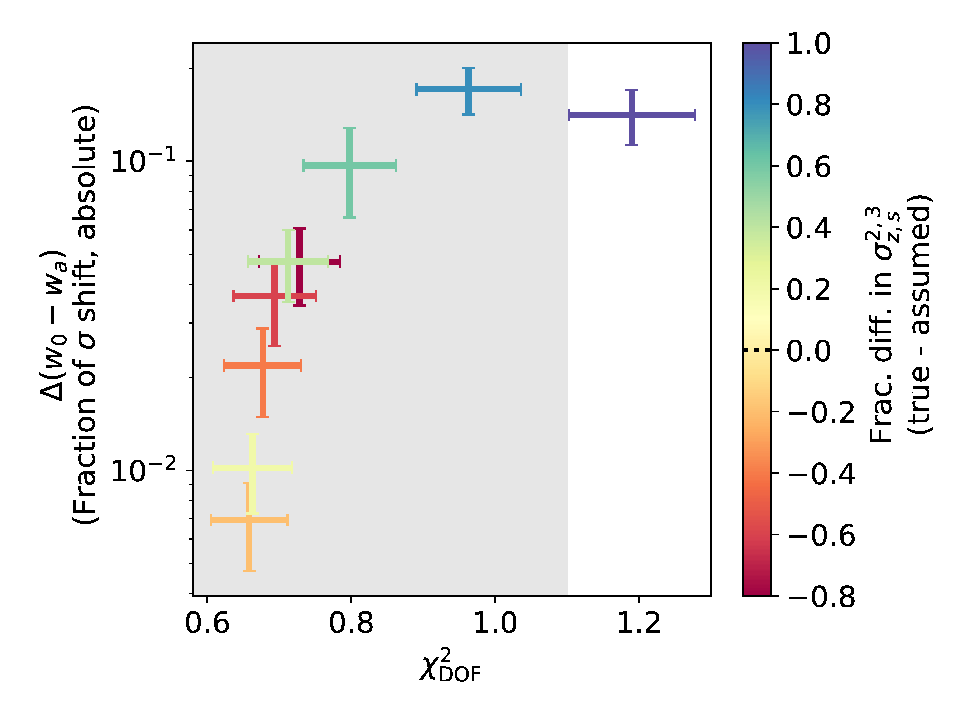}}
\caption{Effect of the mis-specification of $\sigma_{z,s}^{2-3}$ under a truth IA model of TATT with all other parameters non-zero ($a_1$ negative and all others positive) and an assumed IA model of TATT. The assumed, fixed $\sigma_{z,s}^{2-3}$ is $0.05$.  The grey shaded region indicates $\chi^2_{\rm DOF} \le 1.1$. The bias in both $S_8 - \Omega_M$ and $w_0 - w_a$ is below than our threshold of $0.3 \sigma$ for all points, hence no blue shading is included. For further details, see Section \ref{subsubsec:sig23misspec}. }
\label{fig:sig23_misspec_TATT}
\end{figure*}

The results in this scenario are shown in Figure \ref{fig:sig23_misspec_TATT}. The level of cosmological parameter bias for the range of $\sigma_{z,s}^{2-3}$ mis-specification considered is small, and the $\chi_{\rm DOF}^2$ values are largely indicative of a good fit. If we consider biases to the inferred values of non-cosmological parameters, we also see minimal biases away from truth values over the range of $\sigma_{z,s}^{2-3}$ considered, with only $\eta_1$ displaying significant deviations from its truth value at the extreme values of $\sigma_{z,s}^{2-3}$. We surmise that, in this case, marginalization over the 5 TATT parameters is adequate to prevent the mis-specification of $\sigma_{z,s}^{2-3}$ from propagating to cosmological parameter bias via the degeneracy with $b_{\rm TA}$ (similar to in the first case considered in Section \ref{subsubsec:mispsecsig1}, of mis-specification of $\sigma_{z,s}^1$ while marginalizing over TATT, although in that case $\chi^2_{\rm DOF}$ did indicate a modelling issue).

Given this result, we consider then the effect of assuming a model comprising a subspace of TATT where we fix $b_{\rm TA}=1$. TATT with $b_{\rm TA}=1$ fixed has precedent as an analysis model in the literature \citep{troxel2018dark, samuroff2019dark}, and thus this is a realistic scenario worthy of investigation. Concretely, we look at:
\begin{center}
\begin{tabular}{|c|c|c|}
 \hline
 & Truth & Assumed \\
 \hline
IA  & \makecell{TATT  \\ (par set 10)}  & \makecell{TATT  \\ (par set 10) \\ Fix $b_{\rm TA}=1$} \\
 \hline
Photo-z & \makecell{binned shift\\ (baseline pars) \\ $\sigma_{z,s}^{2-3}$ mis-spec} & \makecell{binned shift \\ (baseline pars)} \\
\hline
\end{tabular}
\end{center}
Note that the truth model is the same as above - the only change is that within the assumed model, $b_{\rm TA}$ is fixed to its fiducial value of 1.

\begin{figure*}
\centering
\subfigure{\includegraphics[width=0.45\textwidth]{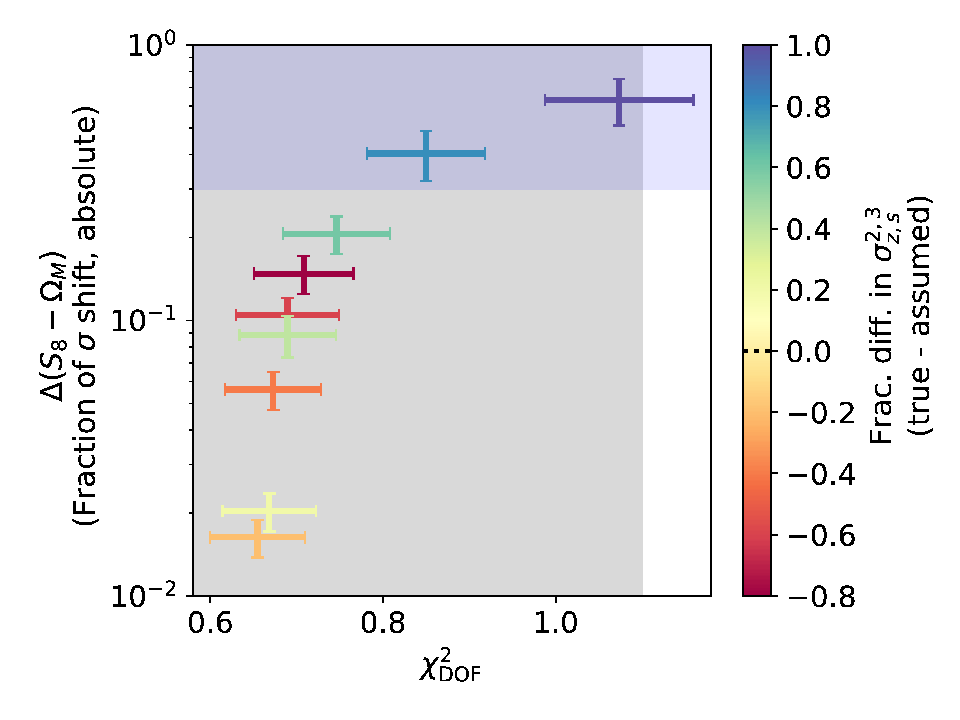}}
\subfigure{\includegraphics[width=0.45\textwidth]{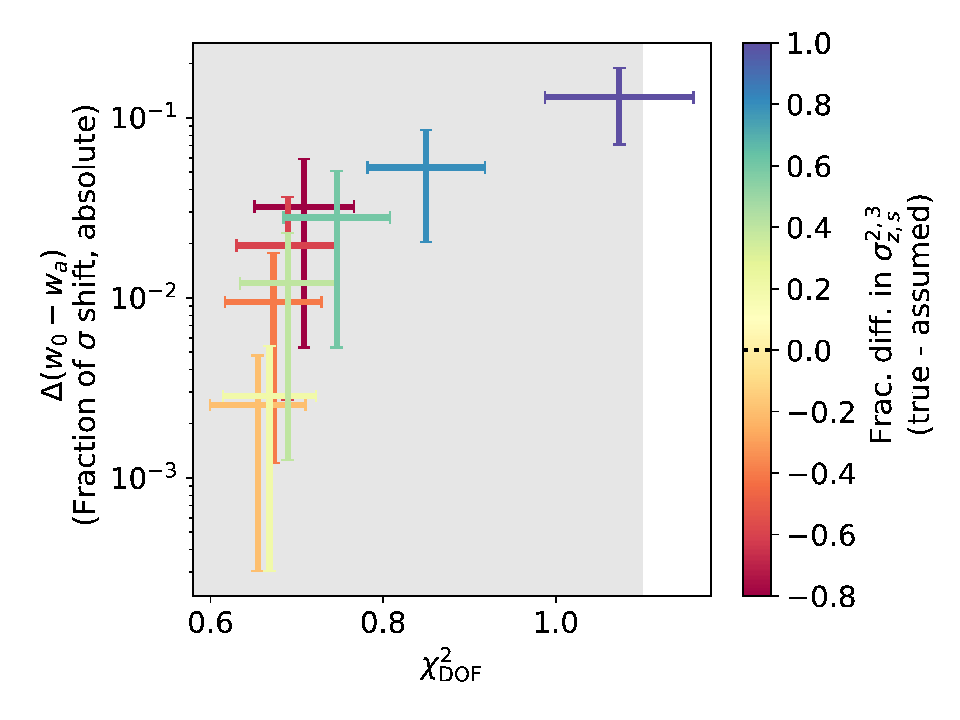}}
\caption{Effect of the mis-specification of $\sigma_{z,s}^{2,3}$ under a truth IA model of TATT with all other parameters non-zero ($a_1$ negative and all others positive) and an assumed IA model of TATT with $b_{\rm TA}$ fixed to its fiducial value of 1. The assumed $\sigma_{z,s}^{2,3}$ is $0.05$. The grey shaded region indicates $\chi^2_{\rm DOF} \le 1.1$. The blue shaded region indicates parameter bias of at least 0.3$\sigma$.  For further details, see Section \ref{subsubsec:sig23misspec}.}
\label{fig:sig23_misspec_TATT_bTAfixed}
\end{figure*}

Results are shown in Figure \ref{fig:sig23_misspec_TATT_bTAfixed}. We see much more severe biases in $S_8 - \Omega_{\rm M}$ than in the above case where $b_{\rm TA}$ is marginalized over, ranging up to greater than $0.5\sigma$. At the same time, the $\chi^2_{\rm DOF}$ values have remained in the same range and in some cases would be indicative of the biased parameter values being a good fit to the data. This is thus the most dangerous regime, where cosmological parameter bias is significant but $\chi^2_{\rm DOF}$ would not provide an empirical diagnostic failsafe. The $w_0 - w_a$ plane, on the other hand, does not see a significant increase in bias from fixing $b_{\rm TA}$. 

\subsubsection{Mis-specification of $b_{\rm TA}$}
\label{subsubsec:bTAmisspec}
Finally, we consider the other type of mis-specification which could interplay with this degeneracy to cause cosmological parameter bias: a mis-specification in $b_{\rm TA}$. We first look at the scenario where our assumed IA model is TATT with fixed $b_{\rm TA}=1$, but the true IA behavior is described by TATT with $b_{\rm TA} \ne 1$:
\begin{center}
\begin{tabular}{|c|c|c|}
 \hline
 & Truth & Assumed \\
 \hline
IA  & \makecell{TATT  \\ (par set 10) \\ $b_{\rm TA}$ mis-spec}  & \makecell{TATT  \\ (par set 10) \\ Fix $b_{\rm TA}=1$} \\
 \hline
Photo-z & \makecell{binned shift \\ (baseline pars)} & \makecell{binned shift \\ (baseline pars)} \\
\hline
\end{tabular}
\end{center}

\begin{figure*}
\centering
\subfigure{\includegraphics[width=0.45\textwidth]{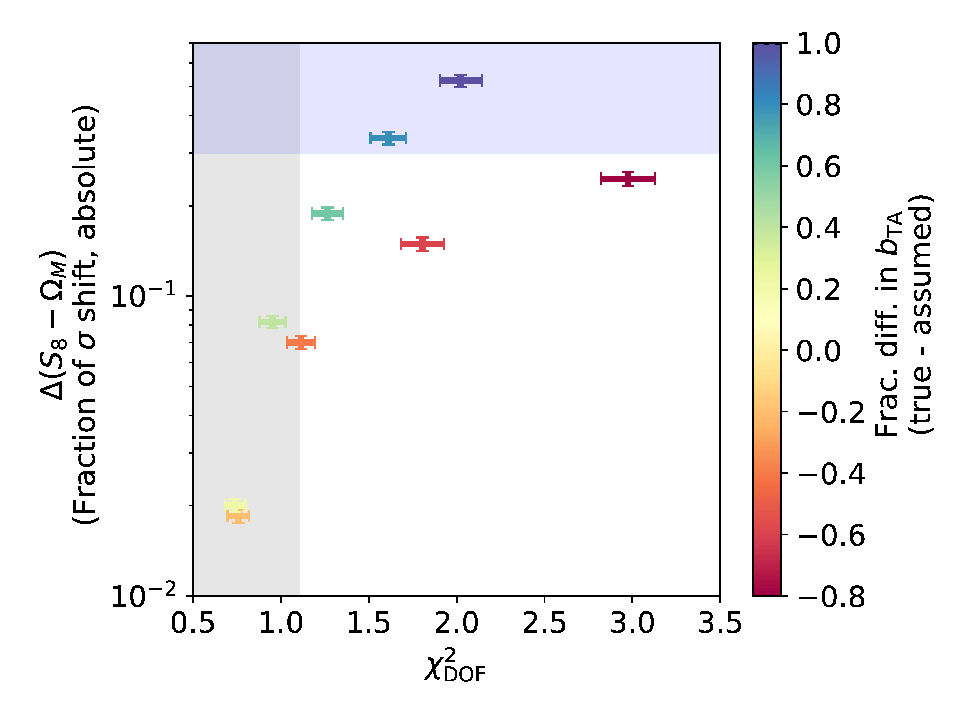}}
\subfigure{\includegraphics[width=0.45\textwidth]{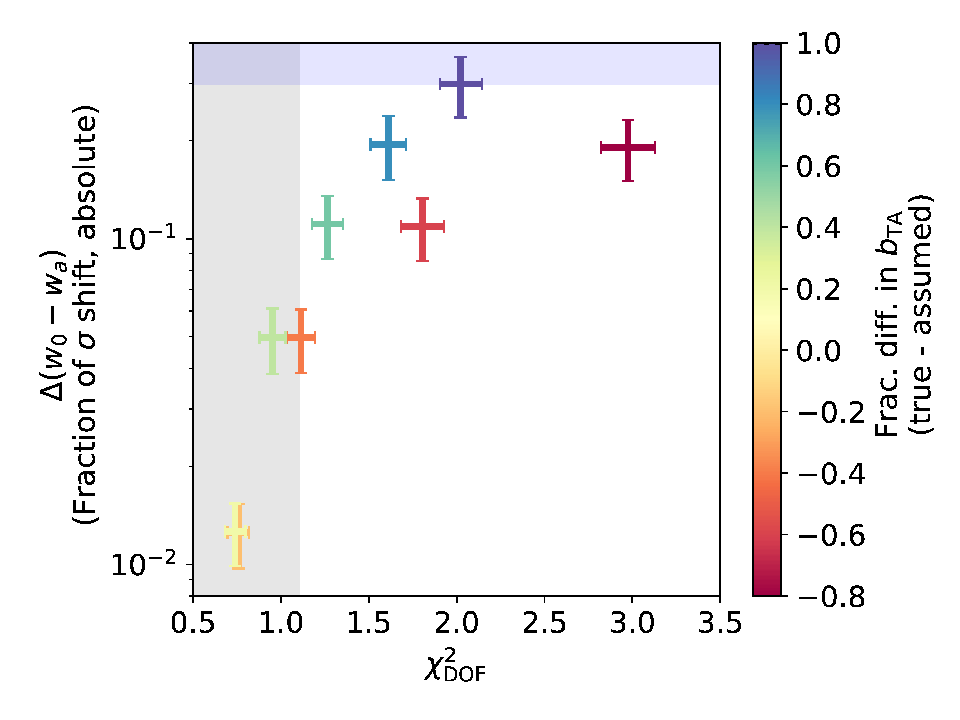}}
\caption{Effect of the mis-specification of $b_{\rm TA}$ under a truth IA model of TATT with all other parameters non-zero ($a_1$ negative and all others positive, IA parameter set 10). The assumed value of $b_{\rm TA}$ is 1. The grey shaded region indicates $\chi^2_{\rm DOF} \le 1.1$. The blue shaded region indicates parameter bias of at least 0.3$\sigma$.  For further details, see Section \ref{subsubsec:bTAmisspec}.}
\label{fig:bTA_misspec_TATT}
\end{figure*}

Results are shown in Figure \ref{fig:bTA_misspec_TATT}. In $\Delta (S_8 - \Omega_{\rm M})$, we see considerable biases in some cases. Of particular interest is the branch corresponding to $b_{\rm TA}^{\rm true}>b_{\rm TA}^{\rm assumed}$. Here, we see that the bias in $S_8 - \Omega_{\rm M}$ rapidly increases with increasing $b_{\rm TA}^{\rm True}$, whilst $\chi^2_{\rm DOF}$ remains near the acceptable good-fit range. Although none of the mis-specifications considered fall into the most dangerous regime ($\Delta(S_8 - \Omega_{\rm M})\ge 0.3\sigma$ and $\chi^2_{\rm DOF} \le 1.1$), the thresholds we have defined are subjective and some mis-specifications fall close to them. The result is a scenario in which, in principle, we could have significant cosmological parameter bias which would not necessarily be identified by a $\chi^2_{\rm DOF}$ diagnostic. 

When $b_{\rm TA}^{\rm true}$ is smaller than the assumed value of 1, the bias in $S_8 - \Omega_{\rm M}$, while still non-negligible, grows less rapidly with absolute deviation of $b_{\rm TA}^{\rm true}$ from $b_{\rm TA}^{\rm assumed}$, while $\chi^2_{\rm DOF}$ scales more strongly. The same behavior is present qualitatively with respect to $\Delta (w_0 - w_a)$, however, with lower absolute values of parameter bias.

We take this consideration a step further by looking at the case where instead of TATT with $b_{\rm TA}$ fixed to 1, we have NLA-z as our assumed model. This set-up involves a persistent mis-specification of $a_2$ and $\eta_2$, as well as the mis-specification to $b_{\rm TA}$. Note that in this case, $b_{\rm TA}^{\rm assumed}$ changes from $1$ above to now $0$ under the NLA-z model assumption.
\begin{center}
\begin{tabular}{|c|c|c|}
 \hline
 & Truth & Assumed \\
 \hline
IA  & \makecell{TATT  \\ (par set 10) \\ $b_{\rm TA}$ mis-spec}  & \makecell{NLA-z \\ (par set 13)} \\
 \hline
Photo-z & \makecell{binned shift \\ (baseline pars)} & \makecell{binned shift \\ (baseline pars)} \\
\hline
\end{tabular}
\end{center}
Our exploration of this scenario reveals that even the case where $b_{\rm TA}^{\rm true} = b_{\rm TA}^{\rm assumed}=0$ results in such significant biases to the $\Delta\bar{z}_s$ parameters and to the power law index $\eta$ of the redshift-dependent NLA model that computed $\chi^2_{\rm DOF}$ values are so high as to be absurd ($O(10^7)$). This is driven by catastrophic shifts to the cosmic-shear cross-spectra involving the lowest-$z$ source bin.  

\section{Discussion and Conclusions}
\label{sec:conc}

As we move rapidly towards early analysis of Stage IV weak lensing and galaxy clustering surveys, having a basis of literature on which to ground our non-cosmological analysis choices and with which to understand their implications is vital for robust analysis and cosmological results. In this work, we have systematically explored the degeneracies between the parameters of intrinsic alignment and photo-z modelling for a 3$\times$2pt analysis of an early Stage IV survey (specifically using an LSST Year 1 like survey as an example), as well as the biases which can result from IA and photo-z model mis-specification. We have focused upon IA and photo-z models which are commonly used in current analysis: TATT for IA and a per-bin mean shift in tomographic redshift bins for photo-z, as well as sub-spaces of these models.

A systematic search for strong degeneracies between IA and photo-z parameters under a representative set of fiducial IA truth models revealed three scenarios of interest. One of these, wherein we see degeneracy between the tidal alignment amplitude parameter $a_1$ and the mean shifts of photo-z modelling, was already well-known in the cosmic-shear literature. We have verified its presence in a 3$\times$2pt scenario and also noted similar degeneracies between the redshift scaling parameter of tidal alignment, $\eta_1$, and both tomographic bin means and variances. 

The other two significant degeneracy scenarios of interest have not to our knowledge been previously explicated; we stress that we have identified them by ensuring that we considered a wide variety of underlying truth IA scenarios rather than assuming a single fiducial IA model consisting e.g. of an NLA-z type behavior. In this way, we identified strong degeneracies between the redshift scaling parameter of the tidal torquing term, $\eta_2$, and the variance of our lowest-$z$ tomographic source bin, $\sigma_{z,s}^1$, in the case where the amplitude of tidal torquing and of source density bias were fiducially non-zero. We also identified a significant degeneracy between the source density bias $b_{\rm TA}$ and several source bin mean and variance parameters, most clearly with $\Delta \bar{z}^1$, $\sigma_{z,s}^{2}$ and $\sigma_{z,s}^3$. This degeneracy was prominent under truth IA models with tidal torquing amplitude and source density bias itself non-zero, but also notably was most prominent under truth IA models where the tidal alignment amplitude $a_1$ took a negative value. 

Having identified these degeneracies, we were able to explore the possible impacts they could have on cosmological parameter inference when combined with relevant model mis-specifications. We find that the most dangerous scenario, in which a model mis-specification causes significant cosmological parameter bias but would not be identified by an empirical goodness-of-fit test, occurs when mis-specifying $\sigma_{z,s}^{2-3}$ under an assumed IA model of TATT with $b_{\rm TA}$ fixed to 1. A mis-specification of $b_{\rm TA}$ under this same model, while marginally outside our subjective `worst case' thresholds, also exhibits this concerning behavior. These types of potential model mis-specification are thus worthy of meticulous care and further consideration, to avoid catastrophic and unidentifiable biases in 3x2pt Stage IV analyses. Other mis-specifications investigated were seen in some cases to induce either severe cosmological parameter bias or $\chi^2_{\rm DOF}$ values significantly above goodness-of-fit thresholds, but not both at once.  

Some general trends were identified in our analysis of the effect of model mis-specifications. Firstly, we find that marginalization over all five TATT parameters is broadly speaking sufficient to negate any serious parameter biases in the $S_8 - \Omega_{\rm M}$ and $w_0 - w_a$ planes, however this does not mean that the inferred values of IA and photo-z parameters will be unbiased under mis-specification, nor does it mean that the resulting $\chi^2_{\rm DOF}$ values will necessarily be indicative of a good fit. As a corollary to this, we see that when mis-specifying a photo-z parameter which is degenerate with an IA parameter, allowing less flexibility in the IA model will tend to produce more bias in the cosmological parameters as the IA parameters will be unable to `absorb' the effect of the mis-specification. We suggest that if using TATT for IA modelling for a fully photometric Stage-IV survey such as LSST, while inferred cosmological parameter values may be robust, it is likely that inferred IA parameter values will be contaminated by residual photo-z errors. Thus, synergistic IA studies with spectroscopic surveys will be important if we seek to achieve unbiased measurement of the IA parameters themselves. 

Secondly, we note a common trend of a lack of symmetry in the effect of a parameter being mis-specified as above or below its true value. In the case of mis-specification of $\sigma_{z,s}^{i}$ parameters, we identify this as being due to the asymmetric integration limits of the lensing kernel, which result in the effective shape of the source redshift distribution being truncated and hence deviating strongly from Gaussian. We see a general trend that it is, perhaps unsurprisingly, worse in terms of our metrics to mis-specify the variance of a tomographic bin as smaller than it truly is than as larger. 

Finally, we see repeatedly that a mis-specification of IA parameters alone, particularly setting $a_2$ and $b_{\rm TA}$ to the moderate and entirely allowed-by-data levels proposed here under an assumed analysis model of redshift-dependent NLA, leads generically to severe cosmological parameter biases in this 3$\times$2pt analysis set-up. This suggests that NLA-z is not likely to be a suitable choice for an IA model for early Stage IV surveys, regardless of any particular interplay between IA and photo-z parameters.

Specifically in the case where we have a mis-specification of $\sigma_{z,s}^1$ when true $a_2$, $\eta_2$, and $b_{\rm TA}$ are non-zero and the model assumed is redshift-dependent NLA, we find that assuming a larger-than-true $\sigma_{z,s}^1$ can compensate neglecting to model the non-zero tidal-torquing and source-density-weighting IA terms. We also see that in a scenario where we mis-specifying $\eta_2$ under a similar full-TATT truth model but an assumed NLA-z model, the degree of mis-specification of $\eta_2$ can have a significant impact on the degree of cosmological parameter bias, and could potentially dominate in a scenario with small but non-zero true $a_2$ and $b_{\rm TA}$, although in the case we consider the mis-specification of  $a_2$ and $b_{\rm TA}$ dominates and the $\chi^2_{\rm DOF}$ would empirically identify the presence of mis-specification.

We have formulated our results in terms of the absolute level of parameter bias in the $S_8 - \Omega_{\rm M}$ (and $w_0 - w_a$) plane, but we can also ask specifically to what extent the mis-specifications that we investigate could be responsible for spuriously shifting the measured value of $S_8$ low, and thus contributing to the $S_8$ tension. We find that for the cases in which the parameter bias in the $S_8 - \Omega_{\rm M}$ plane is above the absolute $0.3\sigma$ threshold, the corresponding shift in $S_8$ is in fact upward from its truth value. Additionally, the Stage IV scenario under consideration represents a step up in constraining power from Stage III, and we would thus expect parameter biases forecast here to be larger (in terms of $\sigma$) than analogous biases in Stage III. The implication is that although our work highlights the possibility of significant biases in $S_8 - \Omega_{\rm M}$ in some Stage IV scenarios, it is unlikely that the mis-specifications under consideration here could explain the $S_8$ tension as seen in current Stage III data. We say this with the caveat that, as discussed above, we consider a relatively simplistic parameterization of photo-z uncertainty; a model which accounts for castastrophic outliers, for example, could yield a different conclusion and is worth further investigation.

Our work builds on previous work (notably that of \cite{fischbacher2023redshift}) in a number of ways, most obviously in that we consider a 3$\times$2pt analysis framework rather than cosmic shear only, and we examine a wider variety of truth scenarios, particularly including those where the redshift-evolution of intrinsic alignment is fiducially non-negligible. We emphasize that we have demonstrated that the sensitivity of 3x2pt cosmological parameter inference to different IA model mis-specifications is highly dependent on the IA truth scenario. This result brings into question the validity of any general claims regarding the sensitivity of cosmological results to IA and photo-z modelling errors. 

There are several clear avenues for possible future work, to extend and improve that which has been presented here. First, it would be of great interest and value to extend the photo-z modelling considered here to be more realistically complex for a Stage IV scenario. Particularly, one could consider the possibility of marginalization over the variance parameters in assumed models, and also include truth-scenario modelling which incorporates catastrophic outliers and other realistic photo-z effects, including more explicit forward modelling of the estimation process for the redshift distribution from photometric information. Secondly, although we have justified the use of Fisher forecasting as our statistical methodology of choice above, it would nevertheless be of interest to extend the results of this work to incorporate MCMC analyses, particularly to capture any effects of non-Gaussian posteriors which are inherently neglected here. This methodological extension could also open up the possibility of exploring whether the mis-specification of photo-z, IA or both may have an impact on model selection metrics not discussed here (e.g. Bayesian evidence), particularly whether the trade-offs between IA and photo-z could conspire to `trick' common model selection metrics into preferring an incorrect cosmological model. Finally, for reasons of concreteness, we have based our forecasting scenario on LSST Year 1, but it would be of interest to examine exactly to what extent our conclusions hold in a Stage IV scenario which more closely follows {\it Euclid}, particularly as the latter expects to have a larger number of tomographic shear bins than we consider here. Although outside the scope of this work, these are all potential avenues for fruitful further exploration. 

\section*{Data Availability}

This article has made use of the data products associated with \cite{mandelbaum2018lsst} which can be accessed at https://zenodo.org/records/5477349. Other data underlying this article will be shared on reasonable request to the corresponding author.

\begin{acknowledgments}
RM was supported in part by a grant from the Simons Foundation (Simons Investigator in Astrophysics, Award ID 620789) and in part by the Department of Energy grant DE-SC0010118. Work at Argonne National Laboratory was supported by the U.S. Department of Energy, Office of High Energy Physics. Argonne, a U.S. Department of Energy Office of Science Laboratory, is operated by UChicago Argonne LLC under contract no. DE-AC02-06CH11357. MMR acknowledges the Laboratory Directed Research and Development (LDRD) funding from Argonne National Laboratory, provided by the Director, Office of Science, of the U.S. Department of Energy under Contract No. DE-AC02-06CH11357. MMR's work at Argonne National Laboratory was also supported under the U.S. Department of Energy contract DE-AC02-06CH11357. For the purpose of open access, the author has applied a Creative Commons Attribution (CC BY) licence to any Author Accepted Manuscript version arising from this submission. 
\end{acknowledgments}


%

\end{document}